\documentclass[fleqn,usenatbib]{mnras}

\usepackage{newtxtext,newtxmath}

\usepackage[T1]{fontenc}

\DeclareRobustCommand{\VAN}[3]{#2}
\let\VANthebibliography\thebibliography
\def\thebibliography{\DeclareRobustCommand{\VAN}[3]{##3}\VANthebibliography}

\usepackage{graphicx}	
\usepackage{amsmath}	
\usepackage{float}

\usepackage[dvipsnames]{xcolor}

\newcommand{\msun}{M$_{\odot} \, $}

\title[Orbital evolution of binary systems]{Orbital evolution of close binary systems: comparing viscous and wind-driven circumbinary disc models}

\author[G.A.~Turpin \& R.P.~Nelson]{
George A. Turpin$^{1}$\thanks{E-mail: g.a.turpin@qmul.ac.uk} \&
Richard P. Nelson,$^{1}$\\
$^{1}$Astronomy Unit, Department of Physics and Astronomy, Queen Mary University of London, London E1 4NS, UK\\
}

\date{Accepted 2024 January 2. Received 2024 January 2; in original form 2023 August 30}

\pubyear{2024}

\begin{document}
\label{firstpage}
\pagerange{\pageref{firstpage}--\pageref{lastpage}}
\maketitle

\begin{abstract}
Previous work has shown that interactions between a central binary system and a circumbinary disc (CBD) can lead to the binary orbit either shrinking or expanding, depending on the properties of the disc. In this work, we perform two-dimensional hydrodynamical simulations of CBDs surrounding equal mass binary systems that are on fixed circular orbits, using the Athena++ code in Cartesian coordinates. Previous studies have focused on discs where viscosity drives angular momentum transport. The aim of this work is to examine how the evolution of a binary system changes when angular momentum is extracted from the disc by a magnetised wind. In this proof-of-concept study, we mimic the effects of a magnetic field by applying an external torque that results in a prescribed radial mass flux through the disc. For three different values of the radial mass flux, we compare how the binary system evolves when the disc is either viscous or wind-driven. In all cases considered, our simulations predict that the binary orbit should shrink faster by a factor of a few when surrounded by a wind-driven circumbinary disc compared to a corresponding viscous circumbinary disc. In-spiral timescales of $\sim 10^6$--$10^7$~yr are obtained for circular binaries surrounded by CBDs with masses typical of protoplanetary discs, indicating that significant orbital shrinkage can occur through binary-disc interactions during  Class I/II pre-main sequence phases.
\end{abstract}


\section{Introduction}

About half of main sequence Sun-like stars in the Solar neighbourhood are members of multiple systems \citep{Abt1983,DM1991}, and their properties provide important hints about their formation and evolution.  Approximately $34\%$ are in binaries, $8\%$ are in triples and $3 \%$ are in higher order systems, and binary orbital periods can be approximated by a log-Normal distribution with a peak at $P_{\rm b}\sim 10^5$ days, corresponding to a separation of $a_{\rm b}\sim 50$~au for an assumed total binary mass of 1.5~\msun \citep{Raghavan2010}. The distributions of binary eccentricities, $e_{\rm b}$, and mass ratios, $q_{\rm b}$, are essentially flat except for an excess of `stellar twins' for $q_{\rm b}>0.95$. \citet{Moe2019} showed the binary fraction of Solar-type stars is strongly anticorrelated with stellar metallicity for periods $< 10^5$~days, but for longer periods the anticorrelation disappears. From this they suggest that close systems form via disc fragmentation, where opacity regulates gravitational instability, and wider systems originate from the fragmentation of optically thin turbulent molecular cloud cores.

Pre-main sequence binary systems are also common \citep{Ghez1993, Leinert1993,Mathieu1994}. Detections of several young embedded systems on close orbits (< 30 au) have been made among Class 0/I sources \citep{Tobin2016}, with some showing evidence for circumbinary material \citep{Tobin2022}. Among Class II sources, several binary systems surrounded by Keplerian circumbinary discs (CBDs) have been detected \citep[e.g.][]{Mathieu1997,Dutrey1994,Czekala2019}. In the case of the short-period spectroscopic binaries AK Sco, V4046 Sgr, DQ Tau and UZ Tau, there is strong evidence that the binary orbits and disc midplanes on large scales are closely aligned with each other \citep{Czekala2019}, such that they might be viewed as precursors to transiting circumbinary planet systems similar to Kepler-16b, -34b and -35b \citep{Doyle2011, Welsh2012}. Recent analyses of close orbiting systems ($a_{\rm b}<10$~au) using spectroscopy from the APOGEE survey indicates that the pre-main sequence binary fraction is consistent with that observed in the field \citep{Kounkel2019}, in agreement with earlier studies \citep{Mathieu1994}. Furthermore, the period distribution of Class II/III binaries in the range 2 - 10,000 days ($a_{\rm b}=0.05$-10~au) is almost identical to the field star population \citep{Kounkel2019}, showing that many of the properties of close binary systems are established in the pre-main sequence phase.

Understanding the formation and early evolution of close binary systems remains an active area of research \citep[see the recent review by][]{Offner2023}. Fragmentation of turbulent molecular cloud cores is expected to produce binaries with separations of $\sim$ few 100~au to $\sim$ few 1000~au \citep{Offner2010}, and fragmentation of massive protostellar discs is expected to occur on length scales $\gtrsim 50$~au \citep{kratter2010}. Hence, the formation of close binary systems with $a_{\rm b}< 10$~au requires substantial migration to occur. \citet{Moe2018} considered the formation of very close binaries ($P_{\rm b} \le 10$~days) via Kozai-Lidov oscillations and dynamical instabilities in triple systems, combined with dissipation through stellar tides, and showed that significant additional dissipation from circumstellar/circumbinary material would be required to produce sufficient numbers of close systems to be in agreement with the observed frequency in main sequence and pre-main sequence systems. Simulations that produce clusters of stars from the fragmentation of turbulent molecular clouds successfully produce close binary systems with separations $< 10$~au through a combination of dynamical interactions in multiple systems, accretion, and dissipation of energy through interactions with circumbinary discs and nearby cloud material \citep{Bate2002,Bate2012}. These simulations evolve for time scales on the order of the free fall time ($\sim 10^5$~yr) and are limited to resolving interactions on scales $\gtrsim1$~au, so the ability of this scenario to produce the shortest period binaries is yet to be demonstrated.

Observations and simulations show that close pre-main sequence binary systems are expected to be surrounded by circumbinary discs, and energy and angular momentum exchange with these discs, leading to orbital shrinkage, is one means by which the closest binary systems could arise from initially more widely spaced systems \citep{LinPapaloizou1979,Artymowicz1991}. However, a number of outstanding questions remain about the outcome of interactions between pre-main sequence binaries and circumbinary discs: during which stage (Class 0/I/II) of the pre-main sequence lifetime does most of the orbital evolution occur?; what role do disc-binary interactions have in shaping the distribution of binary eccentricities?; how does the evolution depend on the initial properties of the central binary and the circumbinary disc? In this paper, we start to address this latter question and examine how disc-binary interactions change when angular momentum transport in the circumbinary disc arises through the launching of a magnetised wind instead of via turbulent viscosity. In this initial proof-of-concept study, we use 2D hydrodynamical simulations combined with a prescribed external torque to mimic the effect of a magnetic torque, similar to that adopted in recent studies of disc-planet interactions \citep{Lega2022,Nelson2023}.

Circumbinary discs are thought to play an important role in the evolution of binary black holes as well as in pre-main sequence stellar systems, and may provide a solution to the `final parsec problem'  \citep{Begelman1980, Milosavljevi2003}. Hence they have been the subject of numerous recent studies. Previous hydrodynamical simulations of CBDs have adopted a range of numerical schemes and codes: SPH \citep{Bonnell&Bate1994a, Bonnell&Bate1994b, Ragusa20217}; polar meshes using finite-difference ZEUS-type codes \citep{PierensNelson2007,Mutter2017, PierensNelson2020,Coleman2022} and energy-conserving finite-volume codes such as PLUTO \citep{ThunKley2017,Miranda2017,Sudarshan2022,Penzlin2022}; unstructured meshes using AREPO \citep{Muñoz2019, Munoz2020}; Lagrangian moving meshes using the Disco code \citep{Duffell2020, D’Orazio2021}; Cartesian meshes using Athena++ and MARA \citep{Shi2012, Moody2019,Tiede2020,Zrake2021}. In this work we use the Athena++ code with refined Cartesian meshes.

Hydrodynamical simulations of CBDs show that a common outcome is the formation of a tidally truncated eccentric inner cavity surrounded by an eccentric, precessing circumbinary disc. Gas accretes through the CBD towards the cavity, and then forms circumstellar discs (CSDs, sometime also referred to as mini-discs) around the central stars after streamers of gas are pulled off the inner edge of the CBD by the individual stars. In general, accretion onto the stars is found to be pulsed due to the orbital phasing associated with the generation of the streamers. A number of studies have adopted relatively thick and viscous discs models ($h=0.1$, where $h\equiv H/r$ is the disc aspect ratio, and $\alpha=0.1$)  due to their rapid equilibration properties \citep[e.g.][]{Miranda2017,Muñoz2019,Moody2019}. For equal mass binaries on fixed circular orbits, it is found that interaction between the stars and the surrounding gas causes the binary orbit to expand rather than shrink, primarily because the CSDs exert a positive torque that is larger in magnitude than the negative torque exerted by the CBD. As the disc thickness is lowered to a value of $h\lesssim 0.05$, however, the binaries are found to in-spiral  \citep{Heath2020,Tiede2020,Dittmann2022,Penzlin2022}. Decreasing the viscosity can also lead to in-spiralling binaries \citep{Penzlin2022}. \citet{Munoz2020} studied circular binaries for a range of mass ratios ($0.1 \le q_{\rm b} \le 1.0$) and found binaries with a mass ratio of $q_{\rm b} \ge 0.3$ lead to expanding binaries, even when reducing the viscosity. Studies have shown the eccentricity of binaries can also play a role in the expansion or contraction of the binary orbit \citep{Muñoz2016,Muñoz2019,Munoz2020,D’Orazio2021,Zrake2021}.

The previously described numerical work has largely been carried out in a regime where angular momentum transport in the CBD and CSDs is assumed to arise because of turbulent viscosity, parameterised through the alpha prescription \citep{Shakura1973}, and presumably originating from the magnetorotational instability (MRI) \citep{Balbus1991}. So far, no work has been carried out to investigate how the interaction changes if angular momentum transport and accretion are instead driven by a magnetised wind \citep[e.g.][]{Bai2013}, as is believed to occur in protoplanetary discs where the ionisation fraction is low. In this paper, we study and compare the two leading theories of angular momentum transport in accretion discs, viscous and wind-driven. A key point is that viscous discs evolve diffusively, whereas mass transport in a wind-driven disc is purely advective, such that we expect there to be significant differences in disc structure between the two cases even when the mass flux being transported through the disc is the same \citep{Lega2022,Nelson2023}. We measure the gravitational and accretional torques to determine whether the binary separation increases or decreases, and at what speed. 

The paper is structured as follows. In Section~\ref{sect:numerical_methods} we outline the numerical methods, basic equations, and physical models. In Section~\ref{sect:torque_calc} we discuss the basic theory required to understand how torques are calculated and how they relate to the migration rate of a binary. In Section~\ref{sect:numerical_visc} we calibrate the numerical viscosity that arises from the use of a Cartesian grid, and in Section~\ref{sect:tiede_comparison} we perform a comparative study with previous work carried out by \citet{Tiede2020}. We present our viscous and wind-driven disc results in Sections~\ref{sect:visc_results} and \ref{sect:wind_results}, respectively. In Section~\ref{sect:comparison} we make a comparison between the results obtained from the two different types of disc models, and we discuss how the morphology of the disc effects the torque in Section~\ref{sect:disc_morphology}. Finally, we summarise the work and draw conclusions in Section~\ref{sect:summary_concl}.

\section{Numerical Method and Problem Setups} \label{sect:numerical_methods}

    In this work we present three sets of simulations. The first uses viscous ring spreading experiments to calibrate the level of numerical viscosity that arises because we simulate rotating flows on a Cartesian mesh. The second involves a suite of simulations of CBDs with different Mach numbers (or equivalently disc aspect ratios) to compare the results of our setup with those of \citet{Tiede2020} and \citet{Penzlin2022}. The third involves a comparison between CBD simulations that adopt different prescriptions for the angular momentum transport mechanism acting in the disc: viscous versus wind-driven. In the following sections we describe the numerical method used to perform these different simulation suites and the initial and boundary conditions that were implemented.
    
\subsection{Basic equations and numerical setup} \label{sect:numerical_setup}

    All simulations presented here were performed using Athena++ \citep{Stone2020} in 2D Cartesian coordinates. Athena++ uses a finite volume Godunov scheme to solve the following vertically integrated equations of hydrodynamics
    \begin{equation}
        \frac{\partial \Sigma}{\partial t} + \nabla \cdot (\Sigma \textbf{v}) = \dot{\Sigma}_{\text{sink}}
        \label{eq:Continuity}
    \end{equation}
    \begin{equation}
        \frac{\partial \Sigma \textbf{v}}{\partial t} + \nabla \cdot (\Sigma \textbf{vv} + P\textbf{I} - \textbf{T}_{\text{vis}}) = \dot{\Sigma}_{\text{sink}} \textbf{v} + \textbf{F}_{g}\text{,}
        \label{eq:Momentum}
    \end{equation}
    where $\Sigma$ is the surface density of the disc, $\textbf{v}$ is the velocity of the gas, $P$ is the gas pressure, $\textbf{T}_{\text{vis}}$ is the viscous stress tensor, and $\textbf{F}_{g}$ is the gravitational force per unit area. The terms $\dot{\Sigma}_{\text{sink}}$ and $\dot{\Sigma}_{\text{sink}} \textbf{v}$ represent the accretion of mass and momentum by sink particles that are implemented in our CBD runs, as described below. The isothermal sound speed is defined according to
    \begin{equation}
        c_{s}^{2} = h^2 |\phi|,
        \label{eq:SoundSpeed}
    \end{equation}
    where $h$ is the local aspect ratio in the disc and $\phi$ is the gravitational potential. The aspect ratio is related to the Mach number, ${\cal M}$, through the expression ${\cal M}=1/h$. By default, Athena++ treats the gas as being adiabatic, and we implement an effective locally isothermal equation of state by imposing instantaneous cooling at each timestep, following \citet{Moody2019}.

     The viscous ring spreading simulations described below are performed with a single central star located at the origin of the coordinate system. When binary stars are implemented they are of equal mass and remain on fixed, circular orbits in all simulations, with a semi-major axis of $a_{\rm b} = 1$. The binary centre of mass is located at the origin of the coordinate system. The gravitational potential of the binary is given by
    \begin{equation}
        \phi = \phi^{(1)} + \phi^{(2)} = - \frac{GM_1}{(r_{1}^2 + r_{s}^2)^{1/2}} - \frac{GM_2}{(r_{2}^2 + r_{\rm s}^2)^{1/2}},
    \end{equation}
    where $G$ is the gravitational constant, the total binary mass $M_{\rm b} = M_1 + M_2 = 1$, $r_1$ and $r_2$ are the distances to each star, and $r_{\rm s}$ is the gravitational softening length, chosen to be $r_{\rm s} = 0.05a_{\rm b}$.
    
    Accretion of mass and momentum onto sink particles surrounding the stellar components is implemented through the $\dot{\Sigma}_{\text{sink}}$ term in eqns~(1) and (2), which is given by
    \begin{equation}
        \dot{\Sigma}_{\text{sink}} = - \frac{\Sigma}{\tau_{\text{sink}}}(e^{ -{r_{1}^{2}} / {2r^{2}_{\text{sink}}}} + e^{ -{r_{2}^{2}} / {2r^{2}_{\text{sink}}}} ),
    \end{equation}
    where $r_{\rm sink} = r_s = 0.05a_{\rm b}$ and $\tau_{\rm sink}^{-1} = 8\Omega_{\rm b}$. This accretion prescription is equivalent to that used by \citet{Tiede2020}.
    
    For all runs involving CBDs, the extent of the domain runs between $x,y \in [-32,32]$. We use a base resolution of $N_x \times N_y = 2400 \times 2400$ cells, but implement a static mesh refinement region, $x_{r}, y_{r} \in [-6.4,6.4]$ where the resolution is increased by a factor of two and which contains the binary and the cavity region of the circumbinary disc.  This equates to a cell size of $0.01\dot{3} a_{\rm b}$ within the refinement region, which is similar to the resolution of 0.0156$a_{\rm b}$ within the cavity used in \citet{Tiede2020}.
    
    Our boundary conditions reinforce the initial conditions at each timestep unless stated otherwise.

\subsection{Setup for numerical viscosity calibration}\label{sect:numerical_visc_setup}

    Before carrying out inviscid wind-driven CBD simulations, it is first important to calculate the numerical viscosity of the setup. We must ensure that we are not being dominated by effects from numerical viscosity, and that the dominant driver of mass accretion through the disc is due to the wind-driven torque. A pressureless, viscously spreading ring follows the analytical solution \citep{Lynden-Bell1974}
    \begin{equation}
        \Sigma_{\rm ring}(\tau,x) = \frac{M_{\rm ring}}{\pi R_0^2} \frac{1}{\tau x^{1/4}}I_{\frac{1}{4}}\left(\frac{2x}{\tau}\right) e^{{(-\frac{1+x^2}{\tau})}},
        \label{sigma_ring}
    \end{equation}
    where $M_{\rm ring}$ is the mass of the disc, $x = R/R_0$ and $\tau = 12 \nu t R_0^{-2}$ are the normalised distance and time quantities respectively, $\nu$ is the kinematic viscosity coefficient and $I_{\frac{1}{4}}$ is the modified Bessel function of order 1/4. The surface density of the ring is initialised with this analytical solution, at time $\tau = 0.018$, and the initial velocity is Keplerian. The aspect ratio is set to $h = 0.005$, which allows the ring to act as a pressureless fluid while maintaining numerical stability. We adopt outflow conditions at the boundaries.

\begin{figure}
    \centering
    \includegraphics[width = 8.25cm, height = 5.5cm]{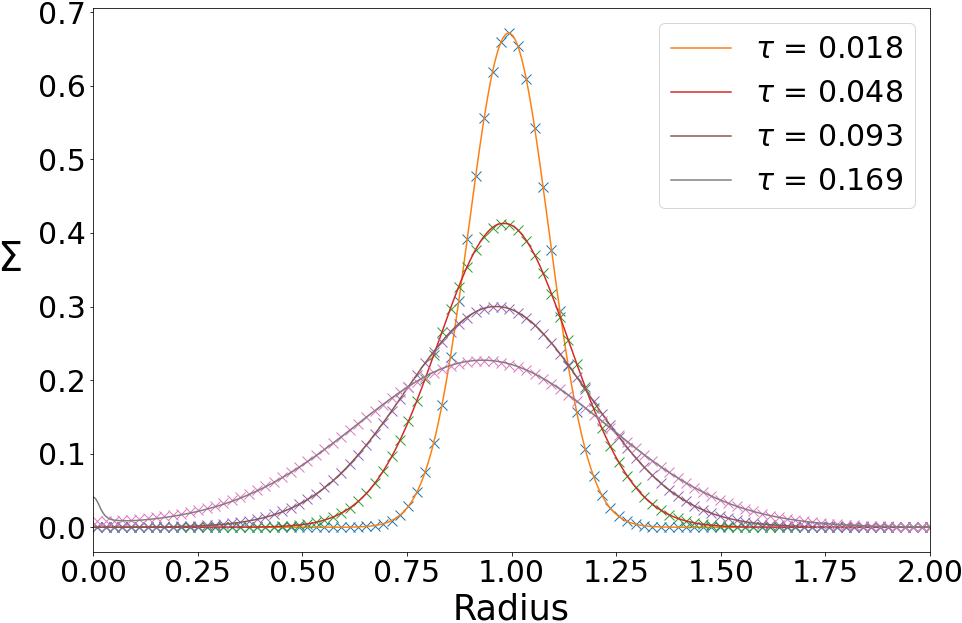}
    \caption{1D surface density evolution of a viscously spreading ring with a kinematic viscosity of $\nu = 1 \times 10^{-5}$. The crosses are the analytical solution at viscous time $\tau$, using Equation \ref{sigma_ring}.}
    \label{visc_nu}
\end{figure}

    We first evolve a viscous ring with a kinematic viscosity of $\nu = 10^{-5}$, to ensure the spreading of the ring correctly follows the analytical solution, and this is shown in Figure \ref{visc_nu}. To calculate the numerical viscosity, we run simulations with $\nu = 0$ for different numbers of grid cells to see how the numerical viscosity varies with numerical resolution. Any spreading of the ring will now be due to the numerical effects, allowing us to calculate the numerical viscosity of the simulations by fitting the results to the analytical solution eqn. ~\eqref{sigma_ring}. Further details of the setup, and and a comparison between the simulation codes PLUTO, FARGO and Athena++, may be found in \citet{Joseph2023}. The results of this viscous ring problem are presented in Section~\ref{sect:numerical_visc}.

\subsection{Setup for Tiede et al. (2020) comparison} \label{sect:tiede_setup}
    In order to calibrate the code against previous simulations of CBDs, we perform simulations of varying Mach number with equivalent initial conditions to \citet{Tiede2020}, allowing us to make a direct comparison to their work.

    The grid setup is as described in Section \ref{sect:numerical_setup}, where the initial condition is a finite disc in quasi-steady state with an inner cavity that has been mildly depleted:
    \begin{equation}
        \Sigma = \Sigma_0 e^{-{(r/r_d - 1)}^2/2} + \Sigma_{\text{floor}}
    \end{equation}
    \begin{equation}
        \textbf{v} = \sqrt{\frac{GMr^2}{(r^2 +r_{\rm s}^2)^{3/2}} + \frac{r}{\Sigma}\frac{\partial P}{\partial R}} \hat{\phi}.
    \end{equation}
    The disc has a peak density at $r_d = 4a_{\rm b}$ and a surface density floor of $\Sigma_{\rm floor} = 10^{-6}$.

    The calculations performed by \citet{Tiede2020} used a constant kinematic viscosity coefficient of $\nu =  \sqrt{2}\times 10^{-3}a_{\rm b}^2\Omega_{\rm b}$, where $\Omega_{\rm b}=1$ is the orbital angular frequency of the binary. The Mach number is varied from 10 to 40 with an interval of 5 (equivalent to varying $h$ between $h=0.1$ and $h=0.025$), resulting in a suite of 7 simulations.
    We present the results from this calibration and comparison in Section~\ref{sect:tiede_comparison}.

\subsection{Setup for wind-driven versus viscous CBDs} \label{sect:wind_setup}

    The grid setup for these runs has been described in Section~\ref{sect:numerical_setup}. To mimic an inviscid disc where angular momentum transport and accretion is due to a magnetically driven wind, we apply an azimuthal acceleration on the gas, which for a disc around a single star would take the form \citep{Lega2022, Nelson2023}

    \begin{equation}
        \label{eq:wind}
        f_{\text{wind}} = \sqrt{\frac{GM_{*}}{R^5}} \frac{\dot{m}}{4\pi\Sigma}.
    \end{equation}
    
    \noindent Here, $\dot{m}$ is the radial flux of gas through the disc, whose value we choose at the start of a simulation and keep constant throughout. This breaking force replicates in a simple manner the torque the disc would feel due to a centrifugally-driven magnetised wind.
    
\begin{figure}
    \centering
    \includegraphics[height = 6.8cm, width = 8.5cm]{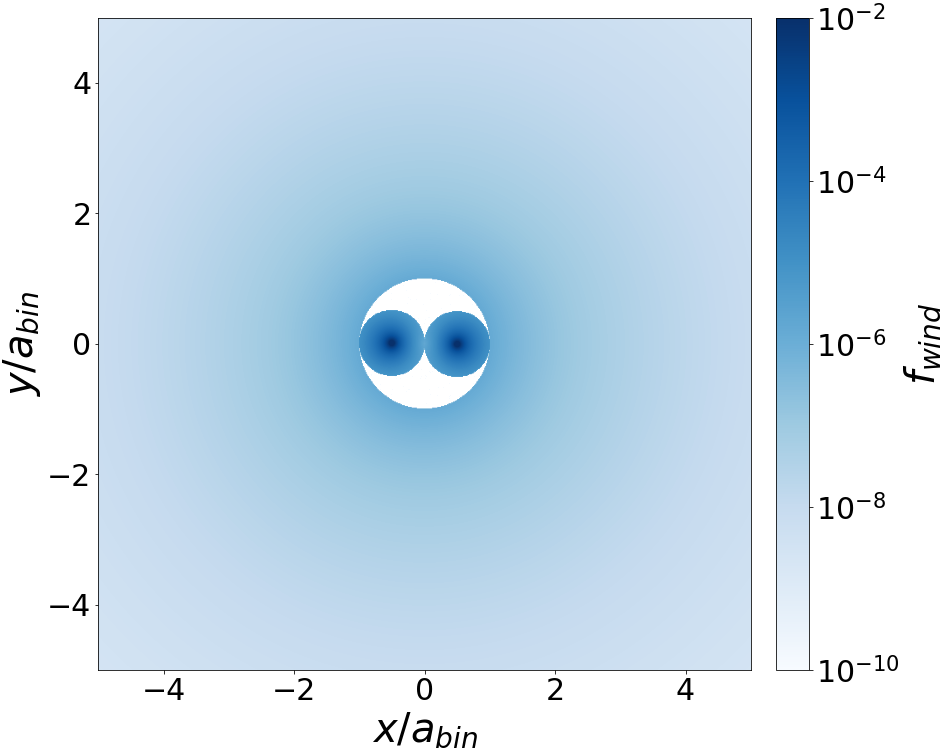}
    \caption{Magnitude of the wind torque on the inner $5a_b$ of the domain, set by the regions defined in Equation \eqref{f_wind}.}
    \label{wind_torque}
\end{figure}

    To adapt this equation for binaries we define three regions of the domain, moving from the outer regions to the inner regions of the CBD-plus-binary system: an outer CBD that orbits around the centre of mass of the binary and in which the radial mass flux is directed towards the centre of mass; an intermediate region that falls inside the cavity and where the gas is treated as being ballistic (we refer to this as the ballistic zone); CSDs/mini-discs that surround each star and within which the radial accretion flow needs to be directed towards the individual stars. We implement the prescribed acceleration as follows:
    \begin{equation} \label{f_wind}
        f_{\rm wind} = 
        \begin{cases}
            \sqrt{\frac{GM_{\rm b}}{R_{\text{COM}}^5}} \frac{\dot{m}}{4\pi\Sigma}, & \text{if $R_{\text{COM}} \geq a_{\rm b}$} \\
            0, & \text{if $R_{\text{COM}} < 1a_{\rm b}$ \&  $R_{1,2} \geq a_{\rm b} $} \\ 
            \sqrt{\frac{GM_{1,2}}{R_{1,2}^5}} \frac{\dot{m}}{4\pi\Sigma}, & \text{if $R_{1,2} \leq 0.5a_{\rm b}$}. \\    
        \end{cases}
    \end{equation}
    In the above, $M_{1,2}$ and $R_{1,2}$ are the mass of and the radius from the individual binary components and $R_{\text{COM}}$ is the radius from the centre of mass of the binary. Figure~\ref{wind_torque} shows how this binary wind torque prescription varies with position in the computational domain. Given that our choice about the size of the ballistic zone is somewhat arbitrary and might affect the results, we have undertaken a study of what happens if different values are adopted. This is presented in appendix~\ref{Sec:Appendix2}, and significant changes in the results are only found to arise when $R_{\text{COM}}  \ge 3 a_{\rm b}$.

    We perform three simulations defined by three values of $\dot{m} = [2.36\times10^{-3}, 6.28\times10^{-4}, 1.23\times10^{-4}]$, where these values assume $\Sigma_0=1$ (which normalises the total disc mass, as defined below). The largest $\dot{m}$ is equivalent to the mass flux rate of a viscous disc with $h=0.05$ and $\alpha  = 0.1$ at $r=1.0$. We then choose the intermediate $\dot{m}$ to be $2\pi\times10^{-4}$ and the lowest value to be 5 times smaller than this. 
    
    In order to study how a wind-driven CBD influences the evolution of the binary orbit compared to a viscously accreting disc, we must also produce equivalent viscous disc models. As $\dot{m}$ is independent of radius in the wind-driven case, we must ensure we have the same in the viscous case. To have a mass flux that is independent of radius for a Keplerian, viscous disc with a power-law surface density profile, we require $\Sigma(R) = \Sigma_0R^{\beta}$ and $\nu(R) = \nu_0R^\delta$, where $\beta = -\delta$. For the viscous models we implement an alpha viscosity prescription, where $\nu = \alpha c_s H$ \citep{Shakura1973}. Since the alpha viscosity profile has $\nu(R) \propto R^{1/2}$, our resulting surface density profile must be $\Sigma(R) = \Sigma_0R^{-1/2}$.
    The alpha values that give rise to the chosen $\dot{m}$ mass flux rates are $\alpha = [0.1, 2.{6} \times 10^{-2}, 5.2 \times 10^{-3}]$. The remaining setup description in this section applies to both viscous and wind-driven disc models.

    We adopt a typical aspect ratio for a protoplanetary disc of $h = 0.05$. The initial velocity profile takes into consideration the required mass flux through the disc due to the wind or equivalent viscosity,
    \begin{equation}
        \textbf{v}_{\phi} = \sqrt{\frac{GMr^2}{(r^2 +r_s^2)^{3/2}} + \frac{r}{\Sigma}\frac{\partial P}{\partial R}} \hat{\phi}
        \label{v_phi}
    \end{equation}
    \begin{equation}
        \textbf{v}_r = -\frac{\dot{m}}{2 \pi r \Sigma_0}.
        \label{v_rad}
    \end{equation}
    
    This ensures the disc does not have to relax under the action of the disc torque and allows a quasi-steady state to be reached more quickly. As the boundaries reinforce the initial conditions, the gas flow into the domain will also have the required radial velocity.
    
    We implement an initial relaxation phase consisting of two stages. Over the first 1000 orbits of each simulation we allow the disc to relax with a single star of mass $M_* = 1.0$ at the centre of the domain. This allows the initial disc profile to relax to the mass flux due to either the wind-driven torque or alpha viscosity. We then introduce a low mass secondary, where mass is linearly redistributed from the initial star to the secondary over the next 500 orbits, to reach the required mass ratio of $M_2/M_1 = 1.0$. During this mass redistribution phase the total mass of the binary is kept constant, with $M_1 + M_2 = 1.0$, the centre of the domain remains the centre of mass of the two stars and the binary semi-major axis remains constant with $a_{\rm b} = 1.0$. We will refer to these 500 orbits as the mass ramping phase, and it is introduced to ensure numerical stability of the low viscosity and inviscid simulations, preventing strong shocks which could lead to negative densities and pressures.

\section{Torques and orbital evolution} \label{sect:torque_calc}
    Before presenting our results, we first outline how the torques on the binary components are calculated and how they relate to the migration of the binary. The analysis below is very similar to that which has been presented previously, for example see \citet{Miranda2017}, \citet{Muñoz2019}, \citet{Dittmann2021} and \citet{Penzlin2022}. We start with the orbital angular momentum of the binary, which can be expressed as
    \begin{equation}
        J_{\rm b} = \mu_{\rm b}\sqrt{GM_{\rm b} a_{\rm b}(1-e_{\rm b}^2)},
        \label{eq:J_b}
    \end{equation}
    where $e_{\rm b}$ is the orbital eccentricity and $\mu_{\rm b}$ is the reduced mass
    \begin{equation}
        \mu_{\rm b} = \frac{q_{\rm b}}{(1+q_{\rm b})^2}M_{\rm b} = \frac{M_1 M_2}{M_1 + M_2}, \; {\rm where} \; q_{\rm b} = \frac{M_2}{M_1}.
        \label{eq:mu_b}
    \end{equation}
    If we consider circular binaries ($e_{\rm b} = 0$), differentiate $J_{\rm b}$ with respect to time and then divide the result by $J_{\rm b}$, we obtain
    \begin{equation}
        \frac{\dot{J_{\rm b}}}{J_{\rm b}} = \frac{\dot{q_{\rm b}}}{q_{\rm b}}\left(\frac{1-q_{\rm b}}{1+q_{\rm b}}\right) + \frac{3}{2}\frac{\dot{M_{\rm b}}}{M_{\rm b}} + \frac{1}{2}\frac{\dot{a_{\rm b}}}{a_{\rm b}}.
    \end{equation}
    
    For equal mass binaries ($q_{\rm b}=1$), the binary migration rate is then given by
    \begin{equation}
        \frac{\dot{a_{\rm b}}}{a_{\rm b}} =
        2 \frac{\dot{J_{\rm b}}}{J_{\rm b}} - 
        3 \frac{\dot{M_{\rm b}}}{M_{\rm b}},
        \label{eq:adot}
    \end{equation}
    showing that a large negative torque, $\dot J_{\rm b}$, and/or a large positive accretion rate, $\dot M_{\rm b}$, lead to rapid shrinkage of the orbit, as expected. We now want to obtain an expression that allows us to determine when the evolution of a binary transitions from out-spiralling to in-spiralling. Rewriting eqn.~(\ref{eq:adot}) as
    \begin{equation}
        \frac{\dot{a_{\rm b}}}{a_{\rm b}} 
        \frac{{M_{\rm b}}}{\dot M_{\rm b}} = 2 \frac{\dot{J_{\rm b}}}{J_{\rm b}} 
        \frac{{M_{\rm b}}}{\dot M_{\rm b}} -3
        \label{eq:adotmdot}
    \end{equation}
    and noting from eqns.~(\ref{eq:J_b}) and (\ref{eq:mu_b}) that
    \begin{equation}
        \frac{\dot{J_{\rm b}}}{J_{\rm b}} 
        \frac{{M_{\rm b}}}{\dot M_{\rm b}} =
        \frac{(1+q_{\rm b})^2}{q_{\rm b}} \frac{1}{\sqrt{G M_{\rm b} a_{\rm b}}} \frac{\dot J_{\rm b}}{\dot M_{\rm b}},
        \label{eq:JdotJ}        
    \end{equation}
    we obtain
    \begin{equation}
        \frac{\dot{a_{\rm b}}}{a_{\rm b}} 
        \frac{{M_{\rm b}}}{\dot M_{\rm b}} = 
        2 \frac{(1+q_{\rm b})^2}{q_{\rm b}} \frac{1}{\sqrt{G M_{\rm b} a_{\rm b}}} \frac{\dot J_{\rm b}}{\dot M_{\rm b}} -3.
        \label{eq:adot_a}
    \end{equation}
    Defining the accretion eigenvalue $\ell_0={\dot J_{\rm b}}/{\dot M}_{\rm b}$ and $\ell=\sqrt{G M_{\rm b} a_{\rm b}}$ in eqn~(\ref{eq:adot_a}), and rearranging terms, gives
    \begin{equation}
        \frac{\dot a_{\rm b}}{a_{\rm b}} = 8 \left(\frac{\ell_0}{\ell} - \frac{3}{8} \right) \frac{\dot M_{b}}{M_{\rm b}}.
        \label{eq:adot_a_final}
    \end{equation}
    Hence, a critical value for the accretion eigenvalue is given by $\ell_0 = 3/8\sqrt{GM_ {\rm b} a_{\rm b}}$, below which $\dot{a_{\rm b}}/a_{\rm b}$ will be negative and the semi-major axis of the binary will shrink.
    
    To calculate the accretion eigenvalue, $\ell_0$, we only require the mass accretion rate and the torque exerted by the surrounding gas on the binary. The total torque on the binary due to the gas has two main contributions, gravitational and accretional torques, $\dot{J_b} = \dot{J}_{\rm grav} + \dot{J}_{\rm acc}$. These two components are defined as
    \begin{equation}
    \begin{aligned}
        &\dot{J}_{\rm grav} = (\textbf{r}^{(1)} \times \textbf{p}_{\rm grav}^{(1)} +  \textbf{r}^{(2)} \times \textbf{p}_{\rm grav}^{(2)})/\Delta t \\
        &\dot{J}_{\rm acc} \text{ } = (\textbf{r}^{(1)} \times \textbf{p}_{\rm acc}^{(1)} +  \textbf{r}^{(2)} \times \textbf{p}_{\rm acc}^{(2)})/\Delta t 
    \end{aligned}  
    \end{equation}
    where,
    \begin{equation}
    \begin{aligned}
        &\textbf{p}_{\rm grav}^{(i)} = \Delta t \int \Sigma \nabla \phi^{(i)} dA \\
        &\textbf{p}_{\rm acc}^{(i)} = \Delta t \int \dot{\Sigma}_{\rm sink}^{(i)} \textbf{v} dA.
    \end{aligned}
    \end{equation}
    The superscripts (1), (2) and $(i)$ in the above refer to the binary components, where $i \in \{1,2\}$.
    The gravitational and accretional torques as well as the mass accretion rate are calculated and saved to file over 600 times per binary orbit, with the accretion eigenvalue being calculated in post production analysis.

\begin{figure}
    \centering
    \includegraphics[width = 8.25cm, height = 5.5cm]{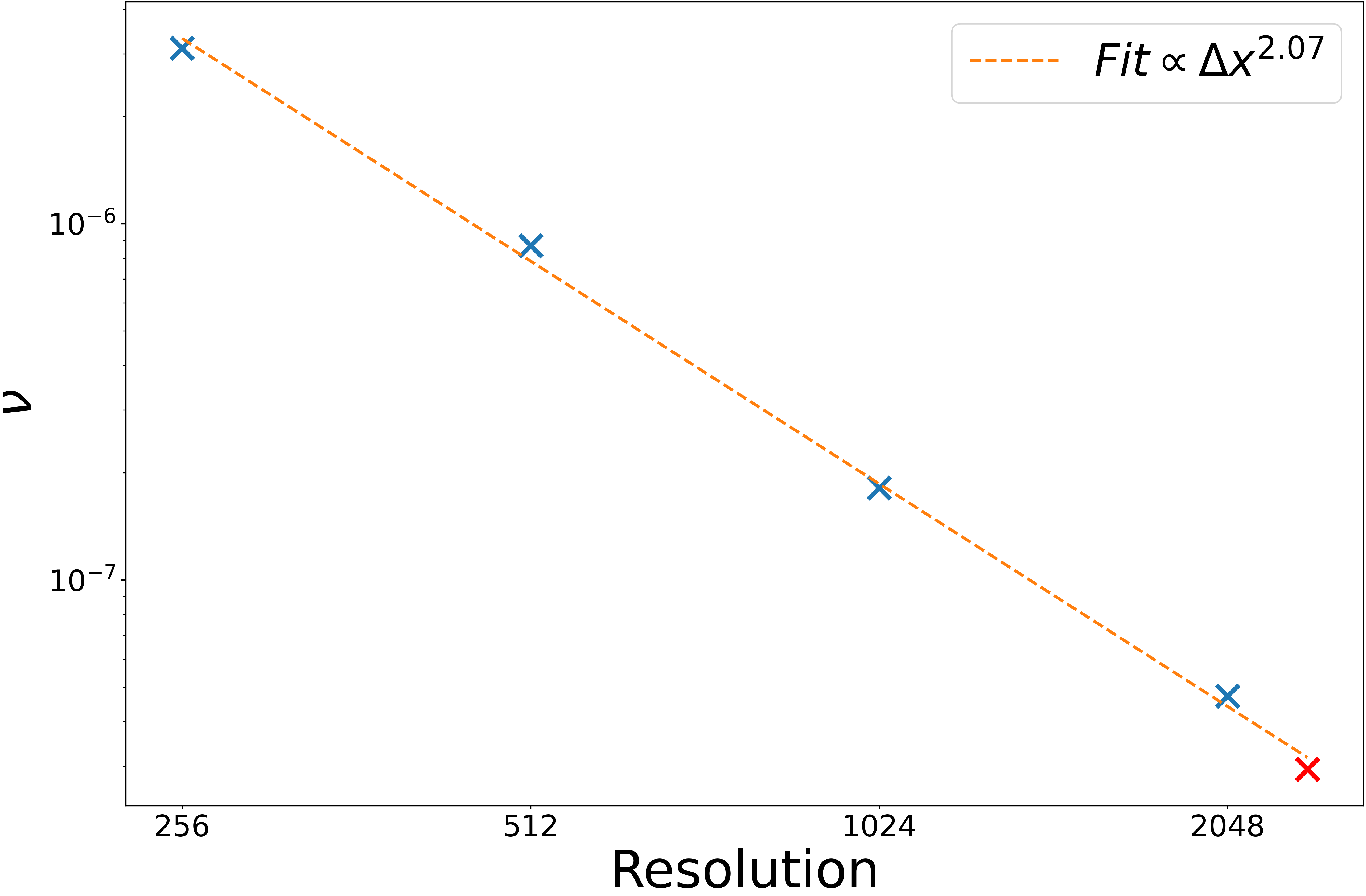}
    \caption{Numerical viscosities for different grid resolutions calculated using Athena++. Values taken from our recent comparison between different simulation codes, presented in \citet{Joseph2023}, are shown as blue crosses. The red cross is the calculated numerical viscosity for the base grid resolution in this work. The fit shows that the numerical viscosity scales as $\nu_{num} = \Delta x^{2.07}$.}
    \label{all_nus}
\end{figure}

\begin{figure*}
    \centering
    \includegraphics[width = 15cm, height = 12cm]{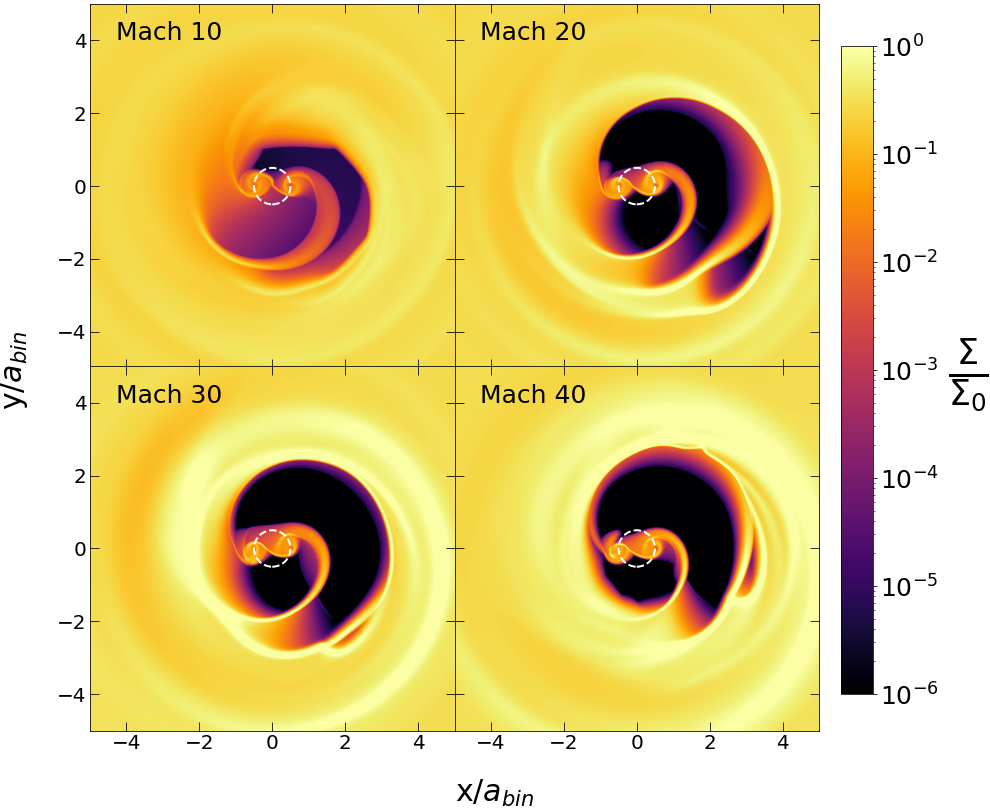}
    \caption{Surface density snapshots after 1000 binary orbits for four values of Mach number. The dashed circle at the centre of the domain represents the binaries orbit. The cavity becomes deeper and the number of spiral density waves increases at higher Mach numbers.}
    \label{tiede_den_plots}
\end{figure*}

\begin{figure}
    \centering
    \includegraphics[width = 8cm, height = 8cm]{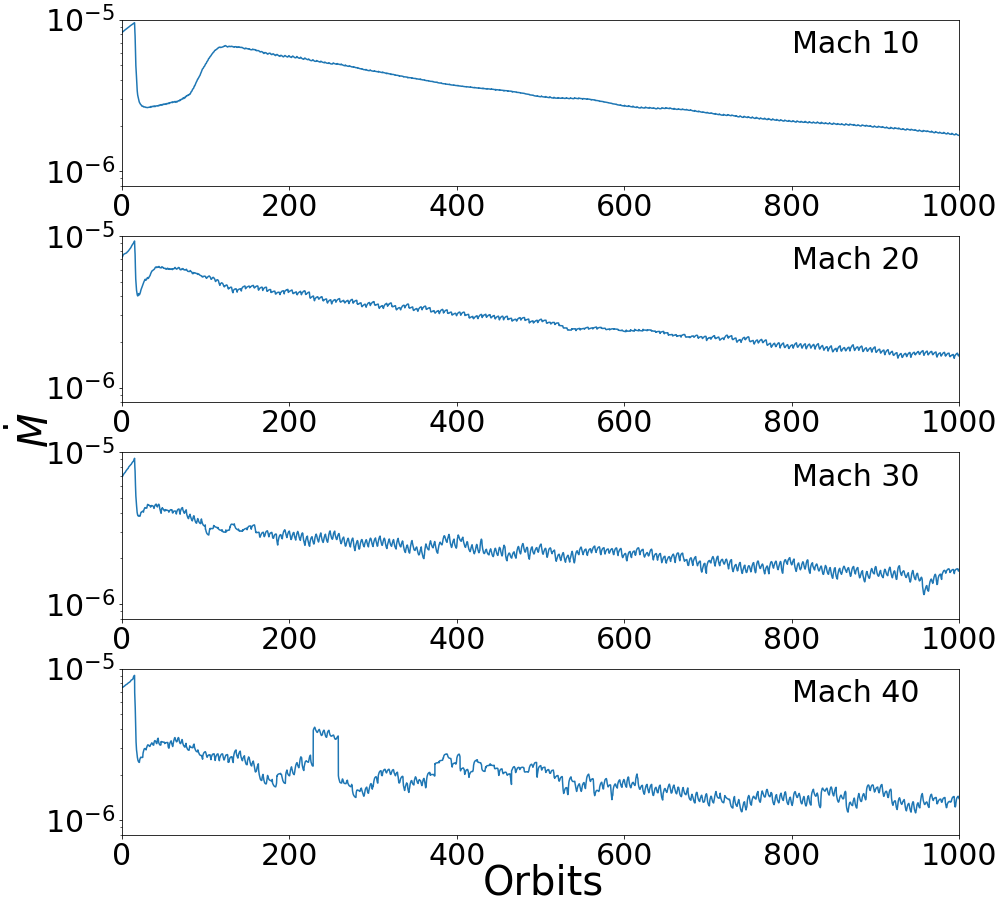}
    \caption{30-orbit averaged mass accretion rate, $\dot{M} = \dot{M}_1 + \dot{M}_2$,  onto the binaries over the whole simulation. $\dot{M}$ becomes slightly suppressed as Mach number increases.}
    \label{tiede_mdot}
\end{figure}

\begin{figure}
    \centering
    \includegraphics[width = 8cm, height = 8cm]{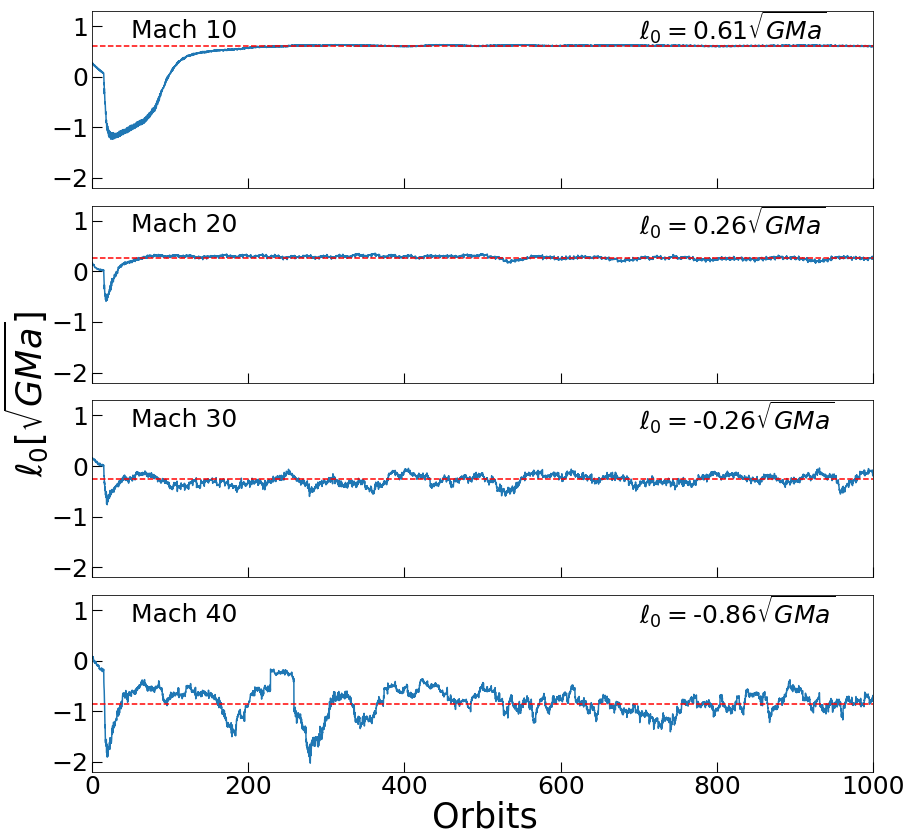}
    \caption{30-orbit averaged accretion eigenvalue, $\ell_0$, over the full 1000 orbits. The dashed horizontal line indicates the value of $\ell_0$ averaged between orbit 500 and 1000. Short term variability increases with Mach number, however there is no secular evolution.}
    \label{30_orbit_l0}
\end{figure}

\begin{figure}
    \centering
    \includegraphics[width = 8.25cm, height = 5.5cm]{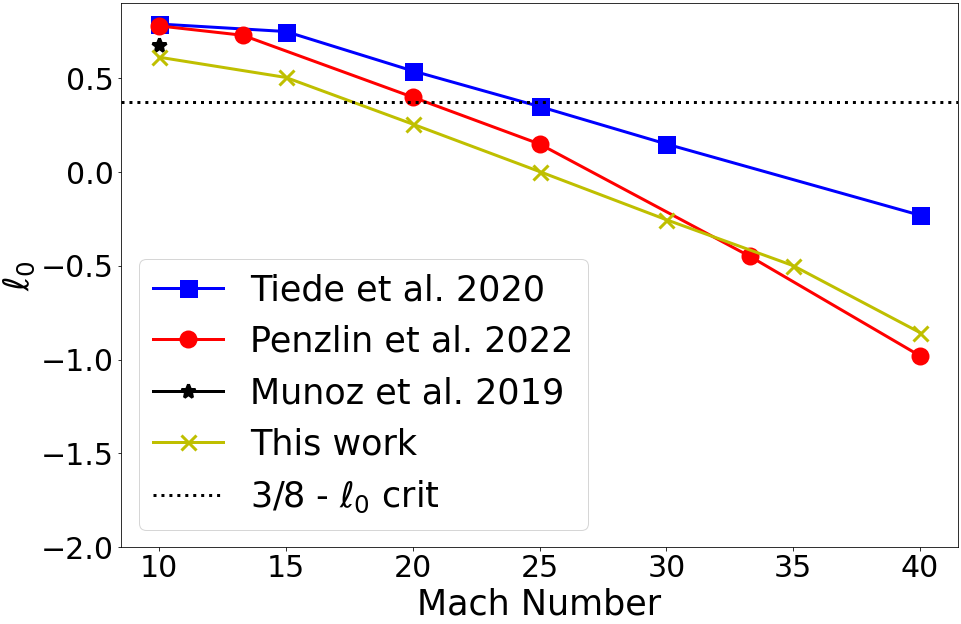}
    \caption{Averaged accretion eigenvalue, $\ell_0$, against Mach number. The horizontal dashed line is the critical value, $\ell_{0_{crit}}$, below which the binaries semi-major axis contract. $\ell_0$ decreases with higher Mach number, crossing the critical value between Mach 15-25. Interactions between the disc and the binary therefore drive the binaries to closer separations for discs with higher Mach number.}
    \label{l0_machno}
\end{figure}

\begin{figure*}
    \centering
    \includegraphics[width = \textwidth]{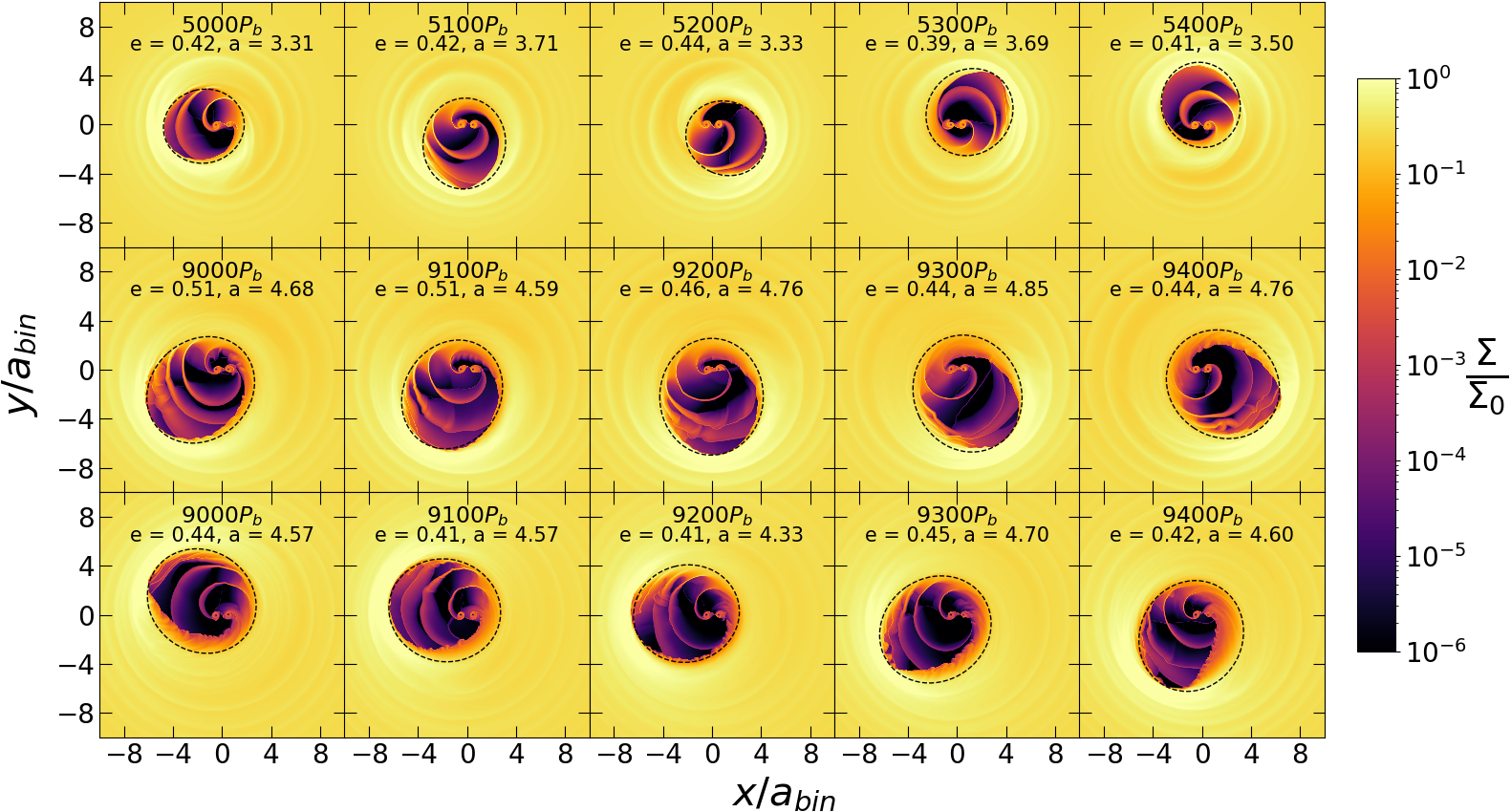}
    \caption{Surface density snapshots for the set of viscous disc simulations. Panels top to bottom descend with mass flux rates, with $\dot{m} =$ 2.36$\times10^{-3}$, 6.28$\times10^{-4}$ and 1.23$\times10^{-4}$.}
    \label{all_alpha_den}
\end{figure*}

\begin{figure}
    \centering
    \includegraphics[width = 8.25cm, height = 5.5cm]{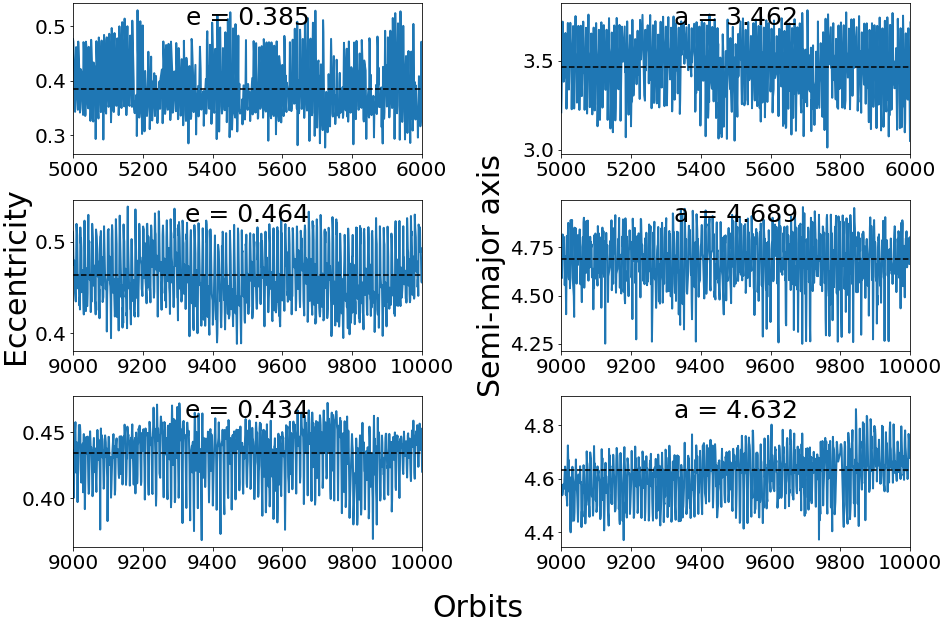}
    \caption{Cavity eccentricity (left) and semi-major axis (right) evolution for the set of viscous disc simulations. From top to bottom the panels show results from the highest, middle, and lowest mass flux rate simulations.}
    \label{e_a_alpha}
\end{figure}

\section{Numerical viscosity study} \label{sect:numerical_visc}
    The results of the ring spreading test for an input value of the kinematic viscosity $\nu=10^{-5}$ is shown in Fig.~\ref{visc_nu}, and it is clear that the simulation follows the evolution predicted by the analytical solution. The results from our suite of inviscid runs performed at different resolutions are shown in Fig.~\ref{all_nus}, where the values of the effective viscosity $\nu$ generated by the numerical diffusion was obtained by fitting the numerical results to eqn.~\ref{sigma_ring} \citep[see also][]{Joseph2023}.
    We report a value for the numerical viscosity of $\nu = 2.9 \times 10^{-8}$ for a resolution $2400 \times 2400$. For a physical disc with realistic parameters ($h \sim$ 0.05), this translates to an $\alpha$ value at $r=1$ of $\alpha = 1.2 \times 10^{-5}$. The numerical viscosity obtained by Athena++ is found to be very similar to that obtained by the PLUTO code using a similar numerical setup \citep{Joseph2023}, resulting in a numerical viscosity that scales as $\nu_{\rm num} \simeq \Delta x^{2.07}$, 
    
    We  note that a resolution of $2400 \times 2400$ is equivalent to the base grid in our domain containing the inviscid, wind-driven CBD, and that the static mesh refinement region has a grid spacing that is smaller by a factor of two compared to the base grid. Using the relationship shown in Fig.~\ref{all_nus}, we estimate the refinement region should have a numerical viscosity of $\nu = 7.5 \times 10^{-9}$. The numerical viscosity in both the base and refinement regions of the domain is therefore small enough to ensure that it is never the dominant driver of gas flow in any of our inviscid, wind-driven disc models.

    The adoption of a cartesian grid means that the global angular momentum in the disc should not be conserved to machine precision. We have examined the level to which the angular momentum is conserved during the highest resolution ring spreading simulation, and we find that the percentage change is just $\sim 4 \times 10^{-5} \%$. Further details are provided in appendix~\ref{Sec:Appendix1}.

\section{Comparison with Tiede et al (2020)} \label{sect:tiede_comparison}
    In this section we present a comparison to the suite of runs performed by \citet{Tiede2020}, which varies Mach number from 10 to 40 for a disc with a constant kinematic viscosity $\nu = \sqrt{2}\times10^{-3}a_b^2\Omega_b$. The setup for these runs is described in Section~\ref{sect:tiede_setup}.

    Figure \ref{tiede_den_plots} shows images of the surface density of the inner $5a_{\rm b}$ of the domain after 1000 binary orbits for four chosen Mach numbers. As Mach number increases, the cavity becomes deeper and we see significantly more nonlinear substructure in the form of spiral density waves at the cavity edge and accretion streamers within the cavity. 

    In Fig.~\ref{tiede_mdot} we plot the total mass accretion rate onto both binary components, $\dot{M}_{\rm b} = \dot{M}_1 + \dot{M}_2$, for the four Mach numbers in which surface density snapshots have been presented. As reported in \citet{Tiede2020}, we also find modest suppression to the mass accretion onto the binary with increasing Mach number. Note these simulations are for finite disc models, and so it is expected that the mass accretion rates will decline with time, as observed.
    
    The time evolution of the accretion eigenvalue, $\ell_0$, is presented in Fig.~\ref{30_orbit_l0} as 30-orbit averaged values, $\langle \ell_0 \rangle_{30} =\langle \dot{J}_{\rm b} \rangle_{30}/ \langle \dot{M}_{\rm b} \rangle_{30}$. Short term variability in $\ell_0$ increases for higher Mach number, however there is no long term evolution. This relationship is also noticed in \citet{Tiede2020}. The dashed line is the value of the accretion eigenvalue averaged between orbits 500 and 1000. We compile these averaged values for all 7 runs in Fig.~\ref{l0_machno}, comparing to the results in \citet{Tiede2020} as well as the results from \citet{Penzlin2022}, who also carried out a similar study using a polar grid instead of a Cartesian one. We also note that \citet{Muñoz2019} performed a simulation equivalent to the Mach 10 run using an irregular adaptive mesh with AREPO. We plot this single result as a star.
    
    Figure \ref{l0_machno} shows that our results are in good agreement with \citet{Muñoz2019} for the Mach 10 case, and as we move to higher Mach numbers our results align with those of \citet{Penzlin2022} while showing the same overall trend with Mach number observed by \citet{Tiede2020}. Due to the highly nonlinear nature of the problem, and how changes in resolution and grid set up can alter the value for $\ell_0$, it is not surprising that different studies using different codes and numerical setups produce somewhat varying results. The dotted line indicates the critical value, $\ell_0 = 3/8 \sqrt{GM_ba_b}$ for which circular, equal mass binaries will have $\dot{a_{\rm b}}/a_{\rm b} = 0$. Comparing all three studies, we see that equal mass binaries are predicted to transition from out-spiralling to in-spiralling within the range of Mach 17-25, for discs with a kinematic viscosity of $\nu = \sqrt{2}\times10^{-3}a_{\rm b}^2\Omega_{\rm b}$ (equivalent to $\alpha = 0.1$ at $r=1a_{\rm b}$).

\begin{figure}
    \centering
    \includegraphics[width = 8cm, height = 8cm]{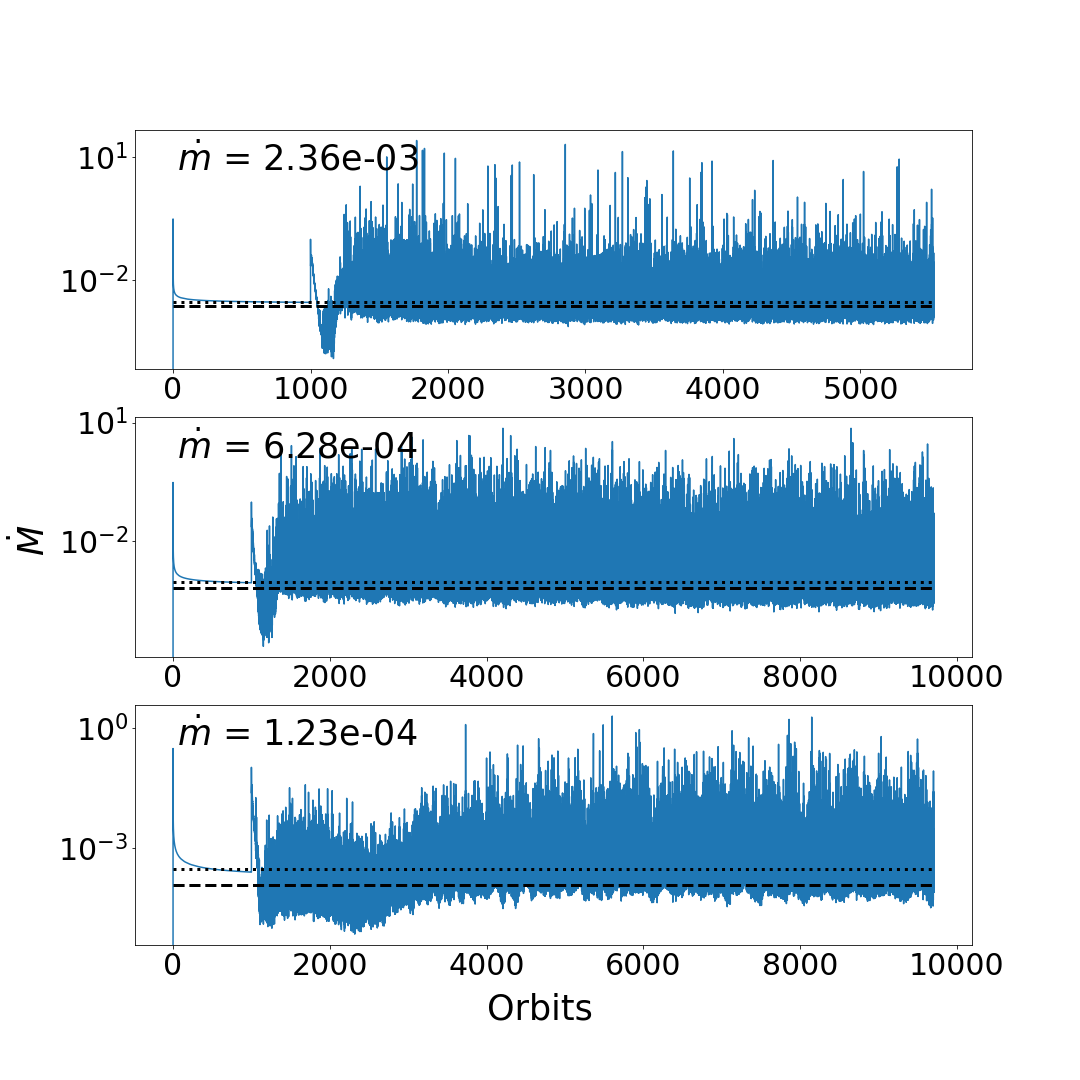}
    \caption{Binary mass accretion rate, $\dot{M_{\rm b}} = \dot{M}_1 + \dot{M}_2$, for each viscous disc simulation. The horizontal dashed line indicates the supplied $\dot{m}$ and the dotted line indicates the $\dot{M}$ into the sinks. The two highest $\dot{m}$ models reach a quasi-steady state almost immediately after the initial 1500 orbit setup period, whereas the lowest $\dot{m}$ model experiences an initial transient phase of approximately 3000 binary orbits.}
    \label{alpha_mdot}
\end{figure}

\begin{figure}
    \centering
    \includegraphics[width = 8cm, height = 8cm]{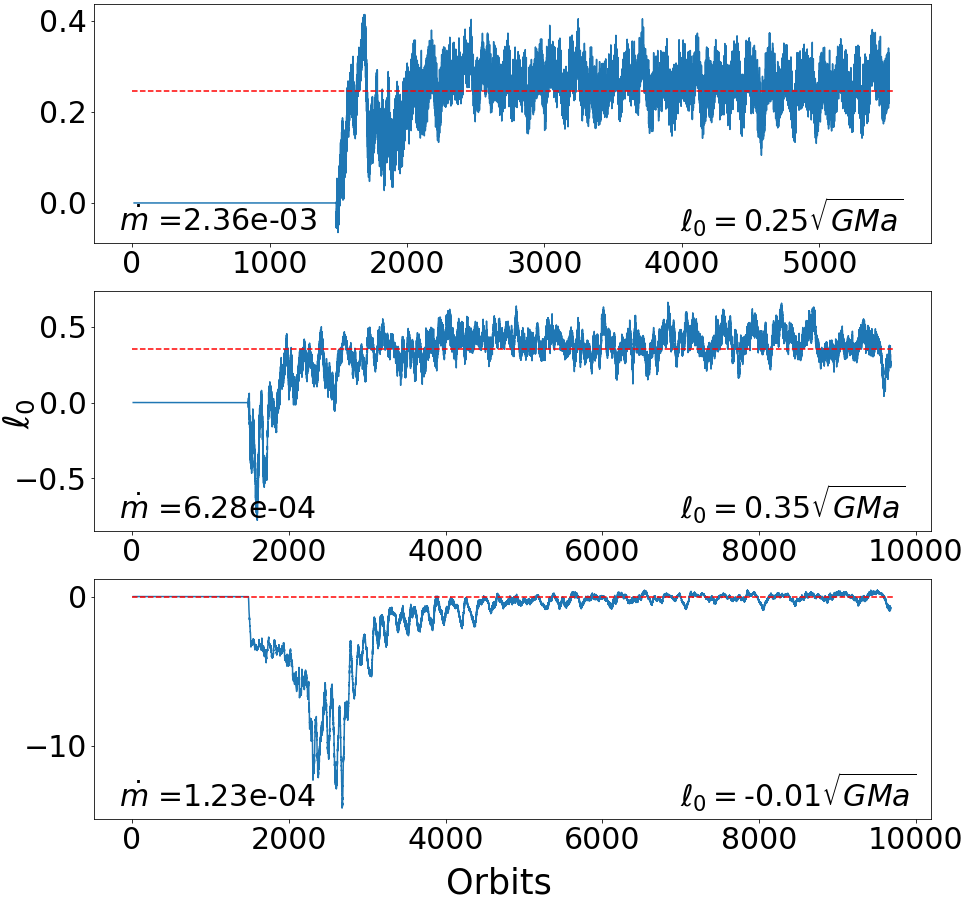}
    \caption{Accretion eigenvalue for the suite of viscous disc simulations. The horizontal dashed line is averaged value over the last 1000 orbits of the simulation. An initial transient phase is seen again in the lowest $\dot{m}$ model.}
    \label{alphas_l0}
\end{figure}

\begin{figure*}
    \centering
    \includegraphics[width = \textwidth, height =9cm]{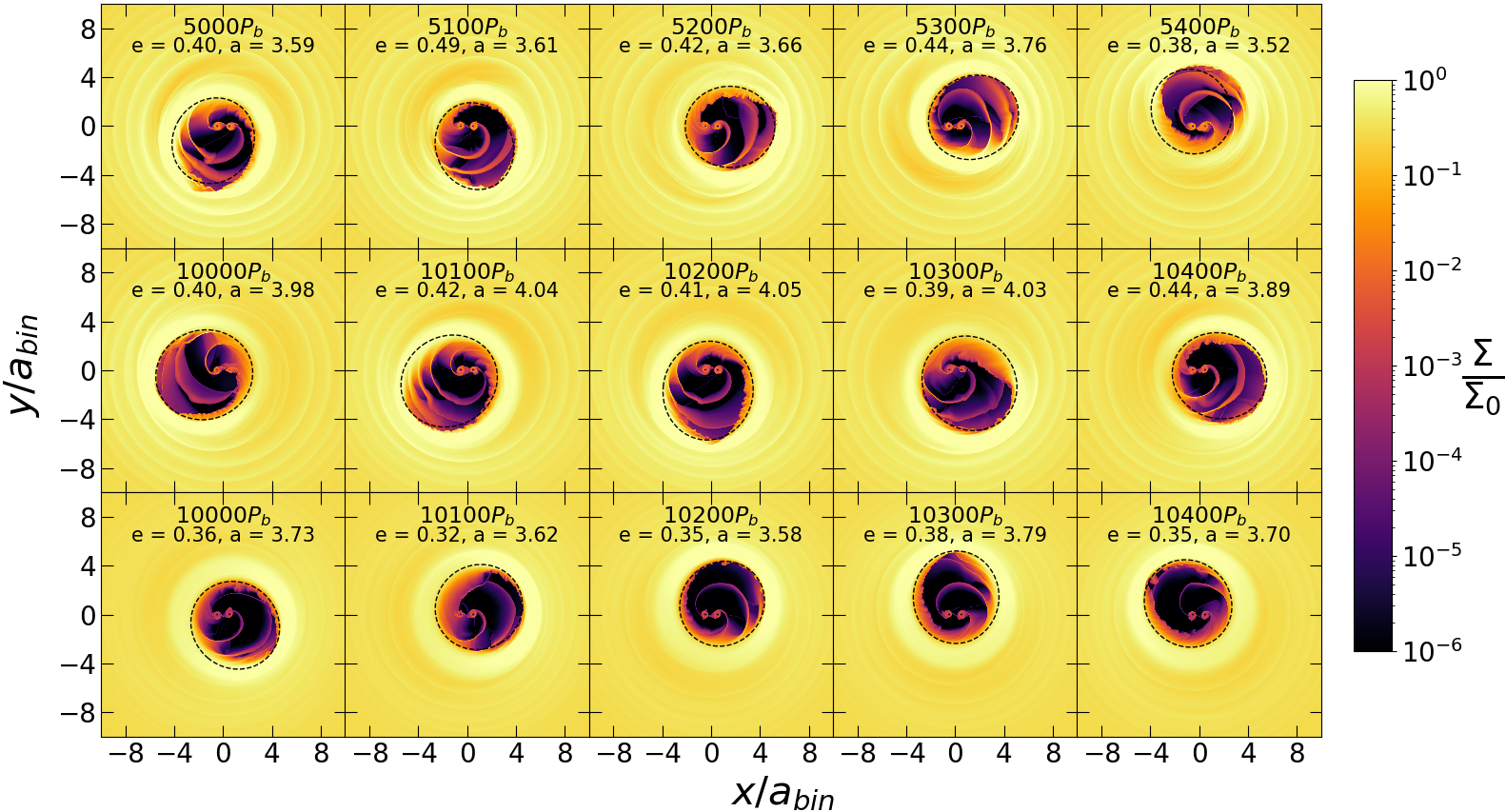}
    \caption{Surface density snapshots for the wind driven simulations. Rows top to bottom correspond to mass flux rates of $\dot{m} =$ 2.36$\times10^{-3}$, 6.28$\times10^{-4}$ and 1.23$\times10^{-4}$ respectively.}
    \label{all_wind_den}
\end{figure*}

\begin{figure}
    \centering
    \includegraphics[width = 8.25cm, height = 5.5cm]{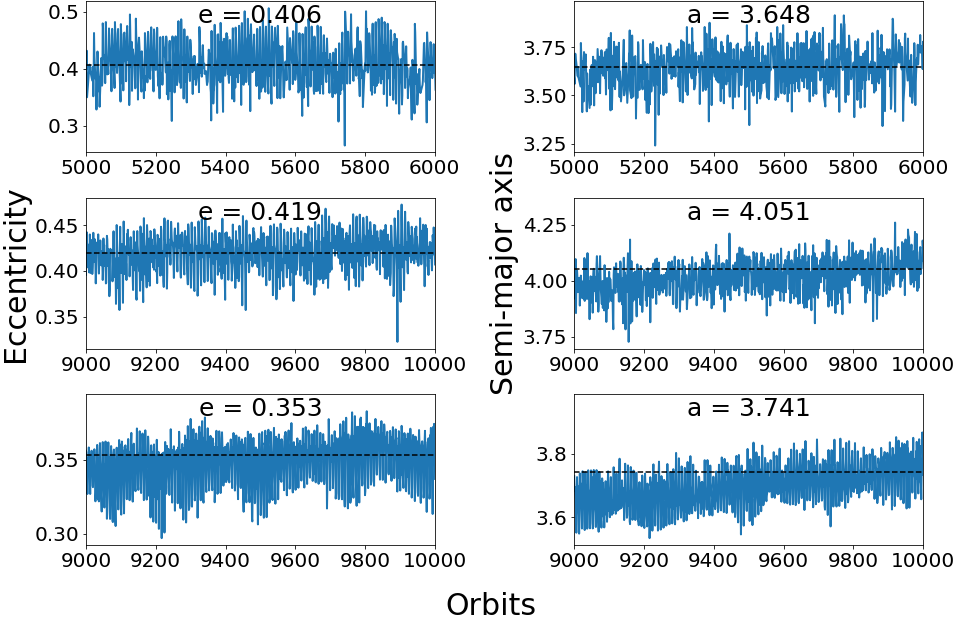}
    \caption{Cavity eccentricity (left) and semi-major axis (right) evolution for the wind-driven disc simulations.}
    \label{e_a_wind}
\end{figure}

\begin{figure}
    \centering
    \includegraphics[width = 8cm, height = 8cm]{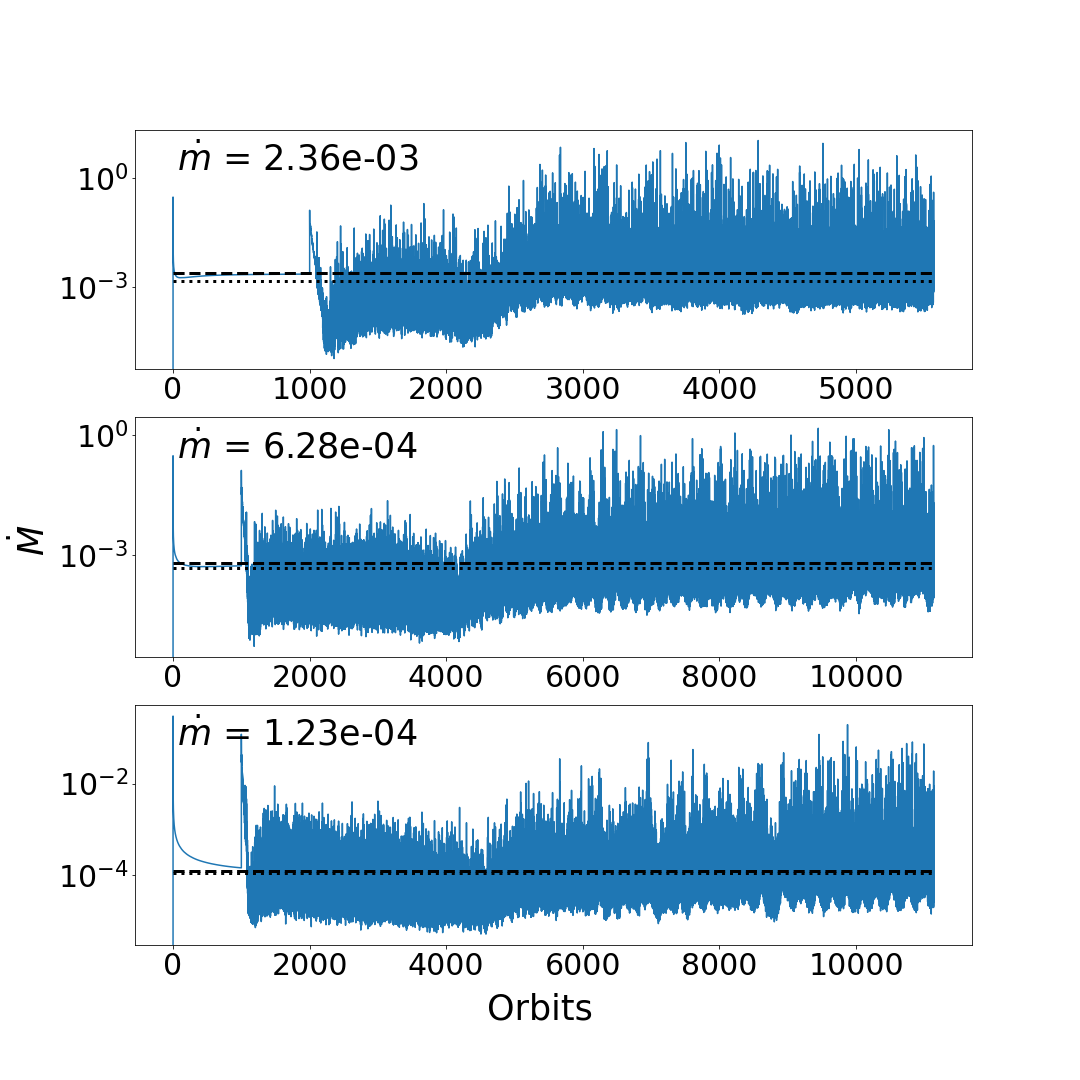}
    \caption{Same as Figure \ref{alpha_mdot} but for the inviscid wind-driven simulations.}
    \label{wind_mdot}
\end{figure}

\begin{figure}
    \centering
    \includegraphics[width = 8cm, height = 8cm]{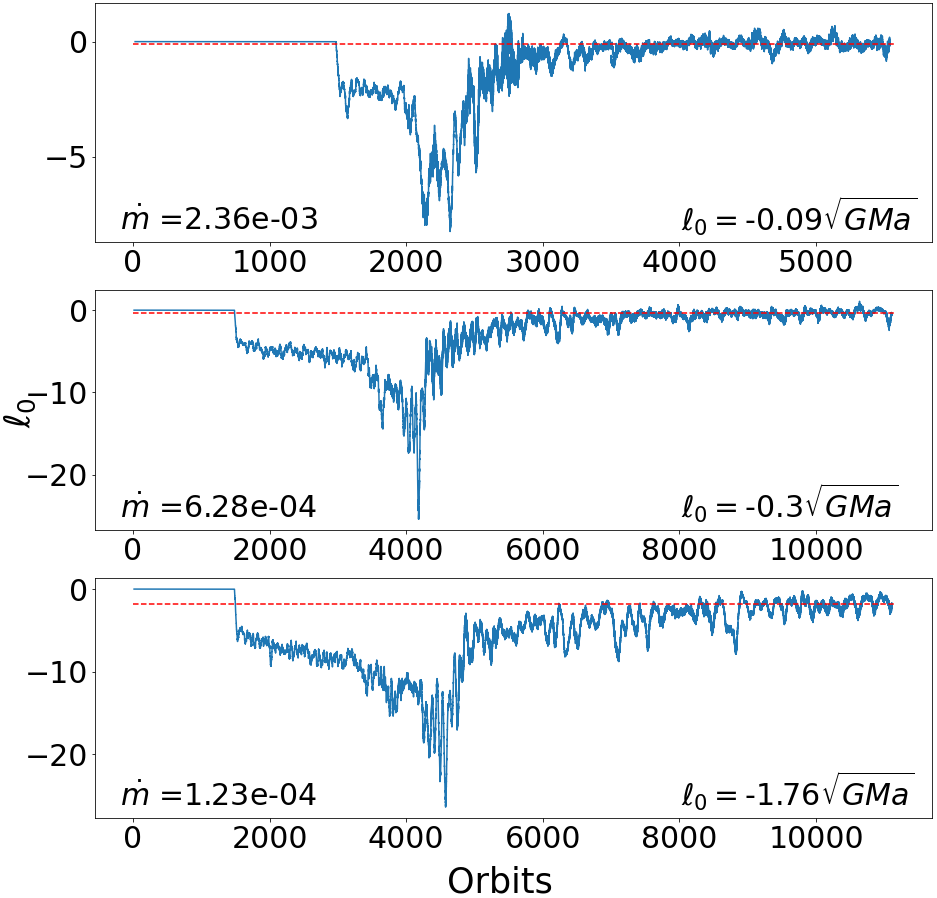}
    \caption{Same as Figure \ref{alphas_l0} but for the inviscid wind-driven disc simulation.}
    \label{wind_l0}
\end{figure}

\section{Viscous disc results} \label{sect:visc_results}
    If we are to answer the question \emph{is an inviscid wind-driven CBD more effective at driving orbital shrinkage of a binary system than a viscous CBD?}, we must first perform simulations in the viscous regime with equivalent mass flux rates through the disc. As described in Section~\ref{sect:wind_setup}, these runs are performed with a Mach number ${\cal M}=20$ (disc aspect ratio $h=0.05$), with three different mass flux rates $\dot{m} = [2.36\times10^{-3}, 6.28\times10^{-4}, 1.23\times10^{-4}]$, which correspond to alpha values of $\alpha = [0.1, 2.6 \times 10^{-2}, 5.2 \times 10^{-3}]$ evaluated at $r=a_{\rm b}$. We present the results from the three alpha viscosity simulations in this section.

    Snapshots of the surface density distributions at different times near the ends of the simulations for each value of the mass flux are shown in Fig.~\ref{all_alpha_den}. As expected, the binary carves out a low-density central cavity region where the semi-major axis of the edge of the cavity is $\sim3-5a_{\rm b}$ \citep{Artymowicz1994}. This central cavity region is asymmetric and contains accretion streamers that transport mass to each binary component. The cavity wall is eccentric in all three simulations, which is expected for circumbinary discs where the central binaries are on circular orbits \citep{PapaloizouNelson2001,Pierens2013,Thun2017}. In the outer disc, the binary excites spiral density waves which propagate outwards. Figure~\ref{e_a_alpha} shows the evolution of the eccentricities and semi-major axes of the cavity edges over the last 1000 binary orbits of each simulation. These two quantities are calculated by drawing a line from the cell with the highest density to the centre of mass. We define the apocentre of the ellipse to be on this line and to coincide with where the density is $10\%$ of the maximum. The line is then extended through the binary centre of mass to the opposite side of the cavity up to the cell with a value of 10\% of the maximum density, which we define as the pericentre of the ellipse. The resulting ellipses are plotted in Fig.~\ref{all_alpha_den}. We see that the run with the largest mass flux has the smallest eccentricity and semi-major axis, the run with the intermediate value of the mass flux has the largest values of eccentricity and semi-major axis, and the run with the smallest value of the mass flux has values of the eccentricity and semi-major axis that are slightly smaller than the intermediate mass flux case. Hence, the behaviour of the cavity is a non-monotonic function of mass flux/viscosity in our runs, although it should be noted that there are large fluctuations in the quantities of interest, and the values for the lowest and intermediate mass flux cases are similar to each other. This relationship contrasts with \citet{Dittmann2022}, who found, for a Mach 20 disc with circular binaries, a monotonic relationship with cavity edge eccentricity and semi-major axis against viscosity. Although we note their simulations used constant kinematic viscosities rather than the alpha prescription.

    Figure \ref{alpha_mdot} plots the mass accretion rate onto the binaries, $\dot{M_{\rm b}} = \dot{M}_1 + \dot{M}_2$, for the full duration of the simulation in each viscous disc case. The panels are defined by their equivalent mass flux rate through the disc, with the highest $\dot{m}$ in the top panel and the lowest in the bottom. The horizontal dashed lines indicate the values of the supplied $\dot{m}$ for each simulation, as described in Section \ref{sect:wind_setup}. The horizontal dotted lines show the total accretion rate onto sink cells averaged over the final 1000 orbits of the simulations. The mass accretion rate in the two highest $\dot{m}$ models reach a quasi-steady state almost immediately after the initial 1500 orbit relaxation and mass ramping period. The lowest $\dot{m}$ model takes almost 4000 binary orbits until reaching an equivalent state where the average mass accretion rate does not appear to exhibit secular evolution. Decreasing the viscosity parameter in turn increases the viscous time-scale (\citet{Lynden-Bell1974}), and so explains why the lowest viscosity model takes nearly 2500 binary orbits more to reach this quasi-steady state than the two larger viscosity models. Comparing the horizontal dashed and dotted lines, we see that the supplied mass flux through the disc and the total accretion rate onto the sinks are very similar to each other in the high and intermediate mass flux runs. Hence, these runs appear to have reached an approximate steady state in which the gas flows through the disc and onto the stars, unimpeded by the tidal torques due to the binaries. In the case with the lowest mass flux, the accretion rate onto the sinks appears to be larger than the mass flow through the disc, suggesting that the disc may be undergoing long term secular evolution because of the smaller viscosity and hence longer viscous time scale in this run. The opening of a gap or cavity can induce a local change in the gas flow rate in a viscous disc, as highlighted recently by \cite{Nelson2023} in the context of a planet opening a gap in a disc, and this possibly explains the behaviour observed in the bottom panel of Fig.~\ref{alpha_mdot}.
    Further discussion of the temporal evolution of the accretion rate onto the  binaries is given in Section~\ref{sect:comparison} below, where we compare the viscous and the wind-driven runs.

    We plot 30-orbit averaged accretion eigenvalues, $\langle \ell_0 \rangle_{30} = \langle \dot{J_{\rm b}} / \dot{M_{\rm b}}\rangle_{30}$, for all three viscous models in Fig.~\ref{alphas_l0}. This averaging is different to the one used in Section \ref{sect:tiede_comparison} in order to be consistent with the work carried out by \citet{Penzlin2022}, with whom we compare results to in Section~\ref{sect:visc_wind_comparison}. The horizontal dashed line is the averaged accretion eigenvalue over the last 1000 binary orbits of the simulation in each case. For the mass flux rates of $\dot{m} = [2.36\times10^{-3}, 6.28\times10^{-4}, 1.23\times10^{-4}]$, we calculate an average accretion eigenvalue of $\ell_0 =$ [0.25, 0.35, -~0.01]$\sqrt{GMa}$ respectively. All of these values are below the critical value, $l_{0_{\rm crit}} = 3/8$, for circular equal mass binaries and therefore correspond to in-spiralling binaries. Initially, there does not appear to be a clear relationship between mass flux and $\ell_0$, however we note that \citet{Penzlin2022} conducted a study comparing different viscosities and their respective $\ell_0$ values across multiple disc aspect ratios and binary mass ratios. Their work found that $\ell_0$ does not vary monotonically with viscosity, and there is a maximum in $\ell_0$ near $\alpha \approx 0.03$. Considering the intermediate mass flux simulation in this work corresponds to an alpha viscosity of $\alpha = 0.026$, we report a similar result.

\begin{figure}
    \centering
    \includegraphics[width = 8.25cm, height = 5.5cm]{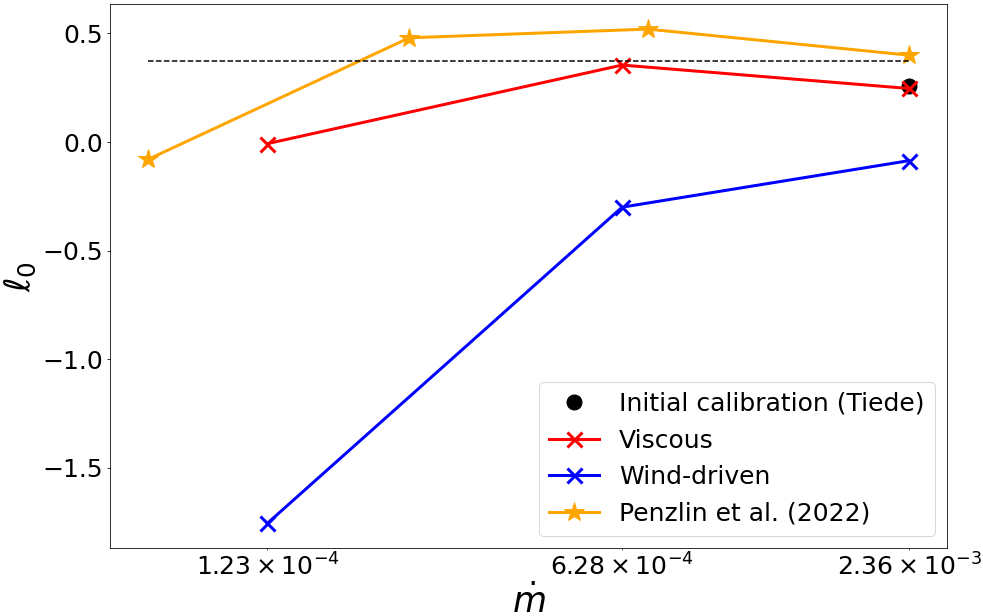}
    \caption{Averaged accretion eigenvalue, $\ell_0$, for all simulations in both the viscous (orange) and inviscid wind-driven (blue) disc regimes. For comparison, results from \citet{Penzlin2022} are plotted in yellow. The solid circle corresponds to the Mach 20 case presented in Section \ref{sect:tiede_comparison}, and the dotted line corresponds to $\ell_0=3/8$. For all mass flux values the inviscid wind model produces an accretion eigenvalue that is significantly less than the viscous counterpart.}
    \label{l0_all}
\end{figure}

\begin{figure}
    \centering
    \includegraphics[width = 8.25cm, height = 5.5cm]{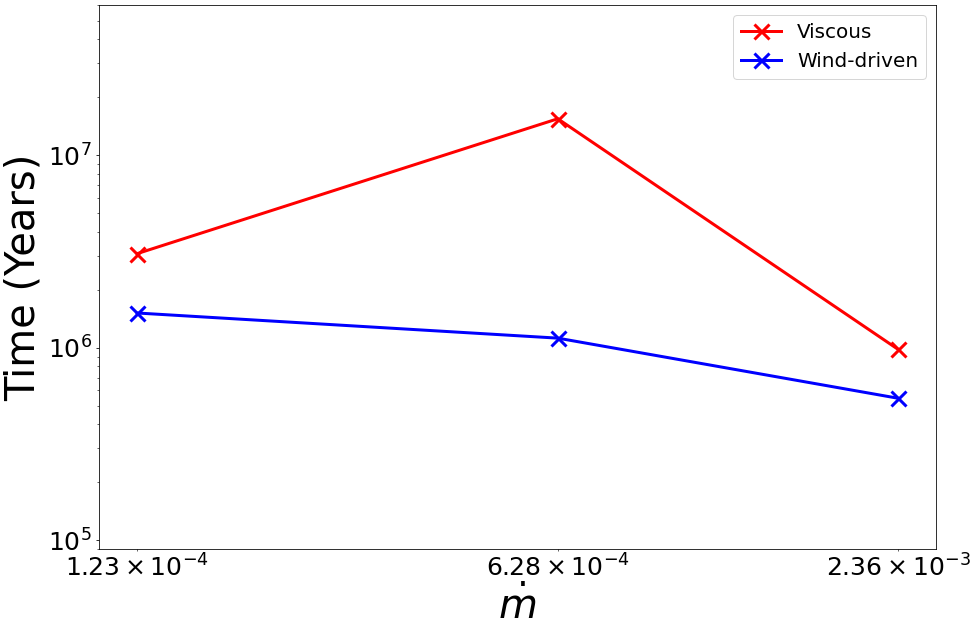}
    \caption{Migration time calculated using the averaged accretion eigenvalues and mass accretion rates in Equation \eqref{eq:adot_a_final}. To calculate these values we have used $M_{\rm b} = M_{\odot}$ and $a_{\rm b} = 1$AU. The disc mass has been scaled to be 5\% of the total binary mass.}
    \label{fig:migration_time}
\end{figure}

\section{Wind-driven disc results} \label{sect:wind_results}

    In this section we present the results from the wind-driven disc models described in Section \ref{sect:wind_setup}, with mass flux rates of $\dot{m} = [2.36\times10^{-3}, 6.28\times10^{-4}, 1.23\times10^{-4}]$. In Fig.~\ref{all_wind_den} we plot surface density snapshots taken near the end of the simulation for each value of $\dot{m}$. Again the binaries open a low-density eccentric cavity region with semi-major axis $\sim 3-5a_{\rm b}$ into which accretion streamers penetrate after being pulled from the cavity wall by the binary components. 
    
    We show the evolution of the eccentricity and semi-major axis of the cavities in Fig.~\ref{e_a_wind}. These values are calculated using the same method described in Section~\ref{sect:visc_results}.  We see that the model with the intermediate value of the mass flux has the largest and most eccentric cavity, whereas the model with the smallest mass flux has the smallest and least eccentric cavity, which is not what is seen with the viscous models. Furthermore, comparing each of the corresponding viscous and wind-driven models, we do not see a clear pattern of behaviour in the structures of the cavities. For the highest mass flux, the wind-driven model produces a wider and more eccentric cavity, whereas the converse is true for the other two mass fluxes we consider. We caution that the lowest right-hand panel in Fig.~\ref{e_a_alpha} and the middle and lowest right-hand panels in Fig.~\ref{e_a_wind} show evidence of residual secular evolution, so it is possible the cavity structures have not fully converged by the end of the runs.

    We plot the total binary mass accretion rates, $\dot{M_{\rm b}} = \dot{M}_1 + \dot{M}_2$, for the three wind-driven simulations in Fig.~\ref{wind_mdot}. As in Fig.~\ref{alpha_mdot}, the panels descend with mass flux rates from top to bottom where the horizontal dashed line shows the supplied $\dot{m}$ through the CBD and the dotted line indicates the total mass accretion rate onto the sink cells. Unlike the viscous runs, a quasi-steady state is not reached immediately in any of the wind driven simulations, instead taking approximately 3000, 5000 and 5000 binary orbits from top to bottom respectively. On longer time scales, however, the mass fluxes through the disc and the mass accretion rates onto the sinks come into decent agreement for each of the runs, indicating that systems are close to being in a steady state where mass flows through the discs and onto the stars at similar rates. This behaviour is quite different to that described by \cite{Nelson2023} for the case of an accreting giant planet in a wind-driven disc. There, the tidal torque of the planet was able to act against the wind-driven torque and effectively halt gas accretion into the gap and onto the planet. In the case of a binary, it appears that the development of a highly dynamic and eccentric disc cavity prevents this mode of evolution from occurring, and instead gas can accrete freely through the cavity and onto the stars. Further discussion of the behaviour of the accretion rates is provided below in Section~\ref{sect:comparison}.
    
    Figure~\ref{wind_l0} plots the 30 orbit averaged accretion eigenvalue for the three simulations. We calculate the averaged value of $\ell_0$ for the three supplied $\dot{m} =$ [2.36$\times10^{-3}$, 6.28$\times10^{-4}$, 1.23$\times10^{-4}$] to be $\ell_0 =$ [-0.09, -0.3, -1.76]$\sqrt{GMa}$ respectively. Again the results indicate binaries that are in-spiralling across all simulations. 
    
    Studying Fig.~\ref{wind_l0} further, we observe the same pattern in all three panels. Each simulations show an initial transient phase where the value of $\ell_0$ is steadily decreasing over the period immediately after the binary mass ramping phase. This is followed by a very sharp drop off and recovery. After the recovery period, the accretion eigenvalue eventually levels out to maintain a steady average. The time taken to reach this final steady state increases for decreasing supplied $\dot{m}$. The level of variability in $\ell_0$ over the course of the simulation also increases for decreasing $\dot{m}$. Further analysis linking the shape of this plot to the morphology of the disc is carried out in Section~\ref{sect:disc_morphology}.

    We leave the key comparisons and analysis between the viscous and wind-driven models to the following section.

\section{Comparison between runs}\label{sect:comparison}

    Here we compile, compare and analyse the accretion rates and the accretion eigenvalue results from both the viscous and wind-driven disc models. We also investigate the contributions to the accretion eigenvalue and the binary orbital migration rate from four regions in the disc. We label these as follows: inner region, ballistic region, cavity region and outer region. The inner region represents the area within $0.5a_{\rm b}$ from each individual binary component, therefore containing the mini-discs/CSDs around each star. The ballistic region covers the gas within $1.0a_{\rm b}$ from the binary centre of mass that is not part of the inner region, corresponding to the area where the wind-driven acceleration is equal to zero, as described in Section~\ref{sect:wind_setup} and indicated in Fig.~\ref{wind_torque}. The cavity region is defined between $1.0a_{\rm b}$ to $2.5a_{\rm b}$ in radius from the binary centre of mass, approximately representing the gas up to the outer edge of the cavity (although note that in the case of an eccentric cavity this may not be strictly true). Finally, the outer region corresponds to gas outside the radius of $2.5a_{\rm b}$ from the binary centre of mass, and therefore contains the main bulk of the outer circumbinary disc. 

\subsection{Viscous and wind-driven disc comparison}\label{sect:visc_wind_comparison}

    Figure \ref{l0_all} compares the accretion eigenvalue averaged over the last 1000 orbits of each simulation for viscous (blue) and wind-driven (orange) simulations. It is clear that a wind-driven accretion disc model produces a significantly lower accretion eigenvalue.
    We also plot results from \citet{Penzlin2022} in Fig.~\ref{l0_all}, who performed similar calculations to our viscous simulations. Although the values for $\ell_0$ in this work are lower than those in \citet{Penzlin2022}, the shape of the plot when moving from higher to lower mass fluxes (or viscosities) remains the same. Both sets of results show a maximum in $\ell_0$ near $\dot{m} = 2\pi \times 10^{-4}$. The slight discrepancy in values between the two results may originate from the different grid setups. \citet{Penzlin2022} carry out simulations on a polar grid with an inner boundary at radius $r=1.0 a_{\rm b}$ and with the binary components orbiting interior to the boundary. The solid circle plotted in Fig.~\ref{l0_all} corresponds to the Mach 20 simulation ($h = 0.05$) we presented in Section~\ref{sect:tiede_comparison} when comparing with \citet{Tiede2020}. Although the initial conditions between the discs differ (finite disc versus power-law surface density profile), the extracted value for the average accretion eigenvalue is almost identical. 

\subsubsection{Binary in-spiral timescales}
    We use the inverse of eqn.~\eqref{eq:adot_a_final} to estimate the time for the binary components to merge in each simulation, and we plot the results in Fig.~\ref{fig:migration_time}. For these calculations we assume an initial separation of 1~AU, a total binary mass of $M_{\rm b} = M_{\odot}$ and a total disc mass of $M_{\rm disc} = 0.05~M_{\rm b}$. 

    The merger times are all in the range $\sim 10^6$--$10^7$~yr, indicating that binaries that remain on near circular orbits will undergo significant orbital shrinkage within the typical lifetimes of protoplanetary discs, which also fall in the range $\sim 10^6$--$10^7$~yr \citep{Haisch2001}. Hence, disc-induced shrinkage of binary orbits from an initially more separated state appears to be a viable mechanism for forming close binary systems during the Class I/II phases of pre-main sequence binaries. In future work, we will consider the evolution of live binaries whose orbits evolve under the direct influence of CBDs,  in order to see if the above result continues to hold when the eccentricities and semi-major axes of binaries evolve in tandem.
    
    We note that a larger-in-magnitude negative value of $\ell_0$ does not always result in faster migration. Equation~\eqref{eq:adot} shows that the net torque on the binary and the mass accretion rate both provide important contributions to the rate of migration. So, although $\ell_0$ can be used to determine whether a binary orbit should expand or contract, its magnitude should not be automatically equated with the migration speed without also considering the mass accretion rate. Larger-in-magnitude and negative $\ell_0$ values can be offset by smaller mass accretion rates (see eqn.~\ref{eq:adot_a_final}). An example of this can be observed when comparing $\ell_0$ (Fig.~\ref{l0_all}) and the migration time (Fig.~\ref{fig:migration_time}) for the smallest $\dot{m}$ wind-driven model. The mass accretion rate in this simulation is smaller than in the runs with larger mass flux values, so even though $\ell_0$ has the most negative value the migration timescale is estimated to be longer than for the other wind-driven disc models.

\subsection{Contributions to $\ell_0$ and ${\dot a}/a$} \label{sect:contributions}
\begin{figure}
    \centering
    \includegraphics[width = 8cm, height = 11cm]{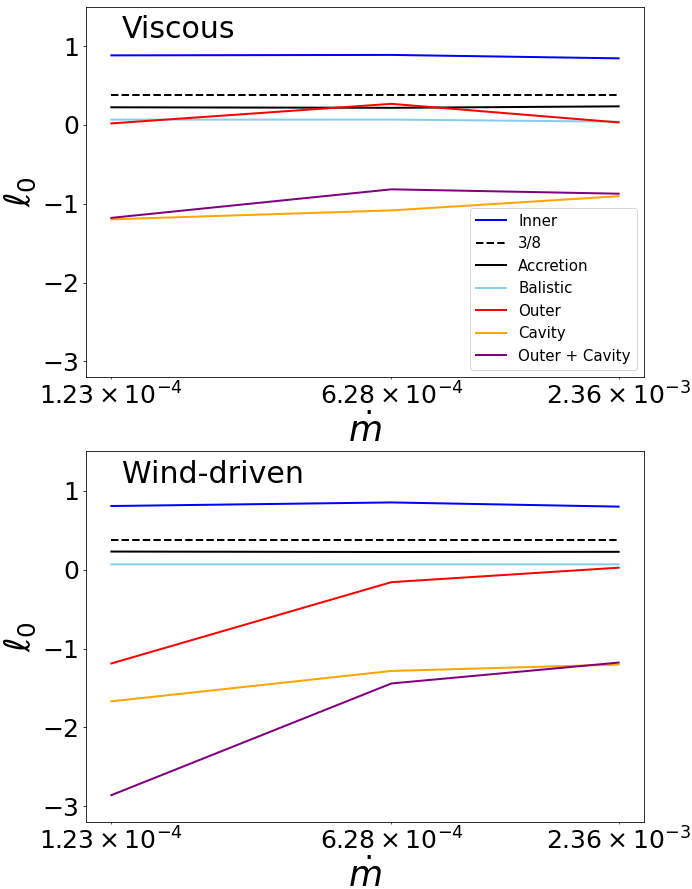}
    \caption{Contributions to the accretion eigenvalue, $\ell_0$, from the regions defined in Section \ref{sect:comparison} for the viscous (top) and the wind-driven (bottom) disc models.}
    \label{l0_all_contribution}
\end{figure}

    Figure \ref{l0_all_contribution} plots the contributions to $\ell_0$ from the different regions of the disc described at the beginning of Section~\ref{sect:contributions} for both the viscous and wind-driven disc runs. The effect of accretion onto the stars is also included, and here we note that the relevant points plotted in Fig.~\ref{l0_all_contribution} combine the effects of both mass accretion and the accretion of momentum. All contributions to $\ell_0$ from accretion are within the the tight range of 0.21 and 0.23 across both sets of simulations, hence providing a contribution below $\ell_{0_{\rm crit}}=3/8$ that would contribute to the in-spiral of a binary, and we discuss the relative contributions from mass and momentum accretion later in this section. There does not appear to be a strong relation between the accretion contribution and mass flux or disc type (viscous or wind-driven).
    
    The inner region contains the mini-discs/CSDs that surround each individual star, defined on the domain by $r_{1,2}<0.5a_{\rm b}$ where $r_{1,2}$ is the distance from each star, respectively. This region has the largest positive contribution to $\ell_0$ and is therefore acting to drive the stars to wider separations. Once again, there is not a strong relationship between mass flux rate and the contribution from the inner regions to $\ell_0$. We find that each wind-driven value is just 5-10\% lower than the viscous counterpart with the same mass flux through the disc although, as we discuss below, this small difference does not necessarily reflect the magnitudes of the torques exerted by the mini-discs because $\ell_0$ also depends on the mass accretion rate onto the binary, which also varies between the runs.
    
    The ballistic zone contains very little gas across all simulations, and therefore has minimal impact on the orbital evolution of the binaries. The contribution to $\ell_0$ due to this small amount of gas across all six runs ranges between values of $\ell_{0_{\rm ballistic}} \approx 0.04 - 0.07$.

\begin{figure}
    \centering
    \includegraphics[width = 8cm, height = 11cm]{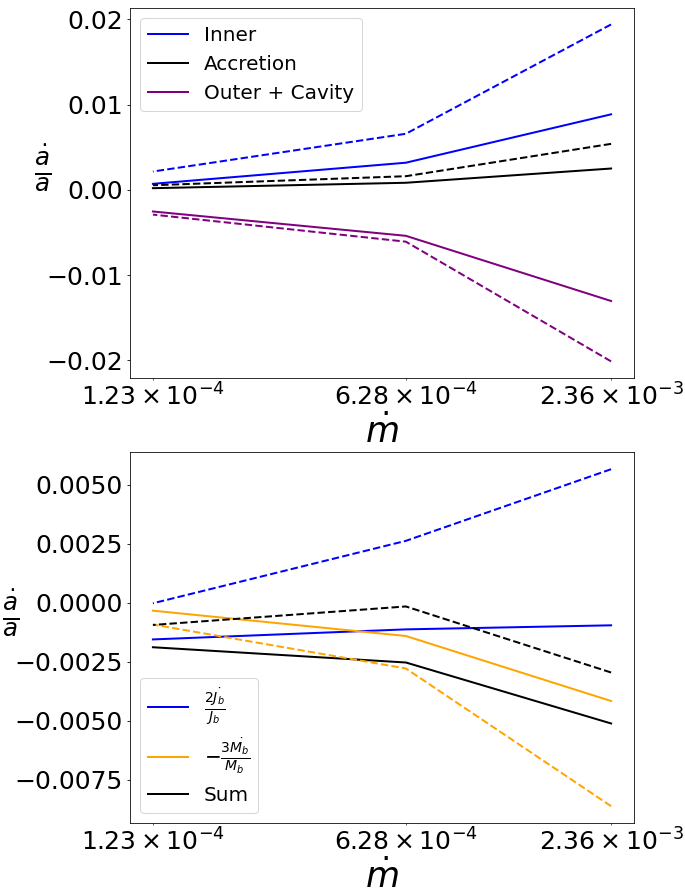}
    \caption{The top panel shows the torque contributions to $\dot a /a$ in eqn.~\ref{eq:adot} from each region of the disc. Dotted lines: viscous disc runs. Solid lines: wind-driven disc runs. The bottom panel shows the total torque contribution to $\dot a/a$, the contribution from mass accretion onto the binary, and the sum of these. Dotted lines: viscous disc runs. Solid lines: wind-driven disc runs }
    \label{fig:adot-contributions}
\end{figure}

\begin{figure*}
    \centering
    \includegraphics[width = 0.8\textwidth]{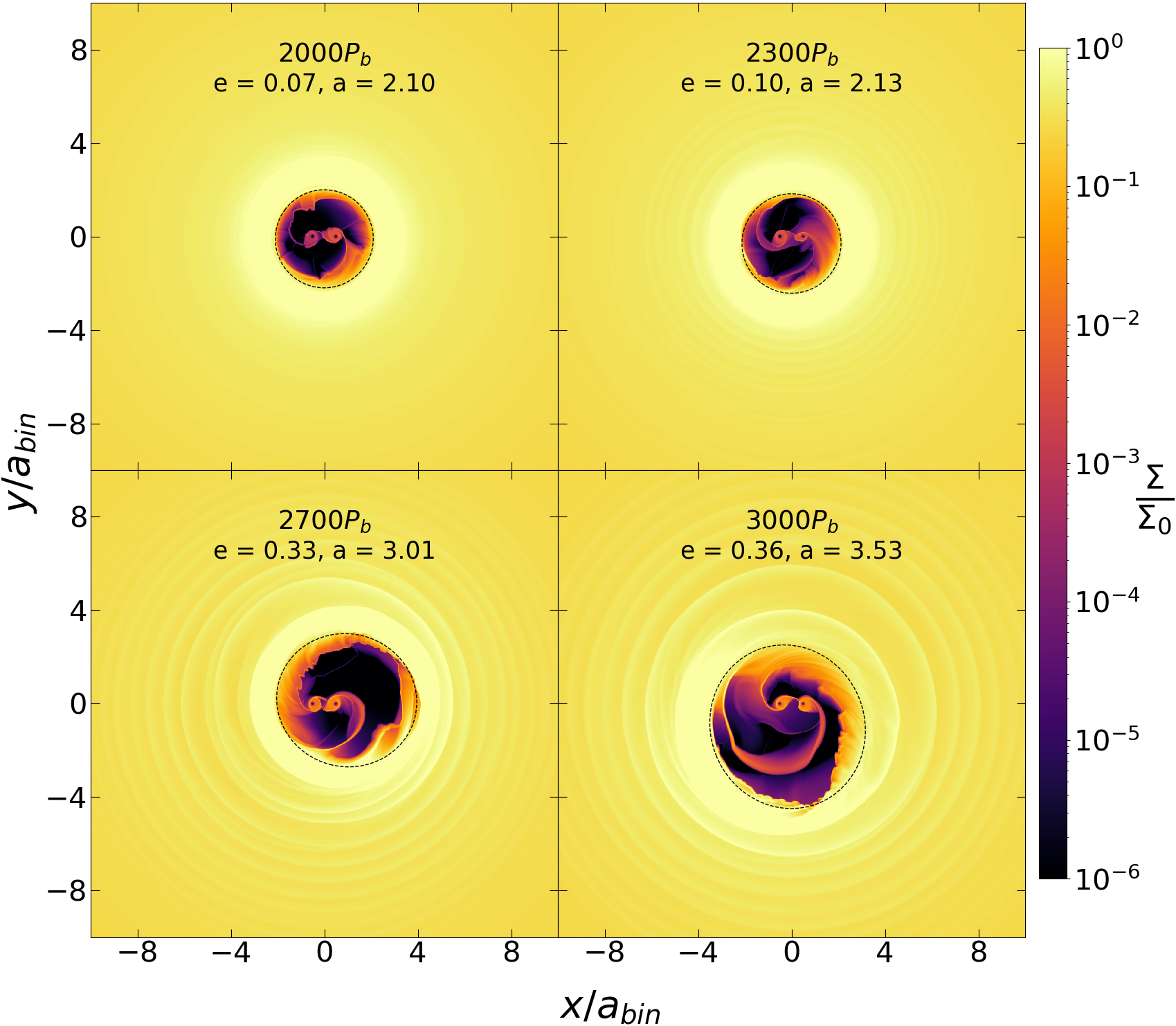}
    \caption{Surface density snapshots for the largest $\dot{m}$ inviscid wind simulation. The circumbinary discs cavity edge can be seen to transition from circular to eccentric. This transition is accompanied by variability in the mass accretion rate and the accretion eigenvalue.}
    \label{surface_den_ctoe}
\end{figure*}

\begin{figure*}
    \centering
    \includegraphics[width = 0.8\textwidth]{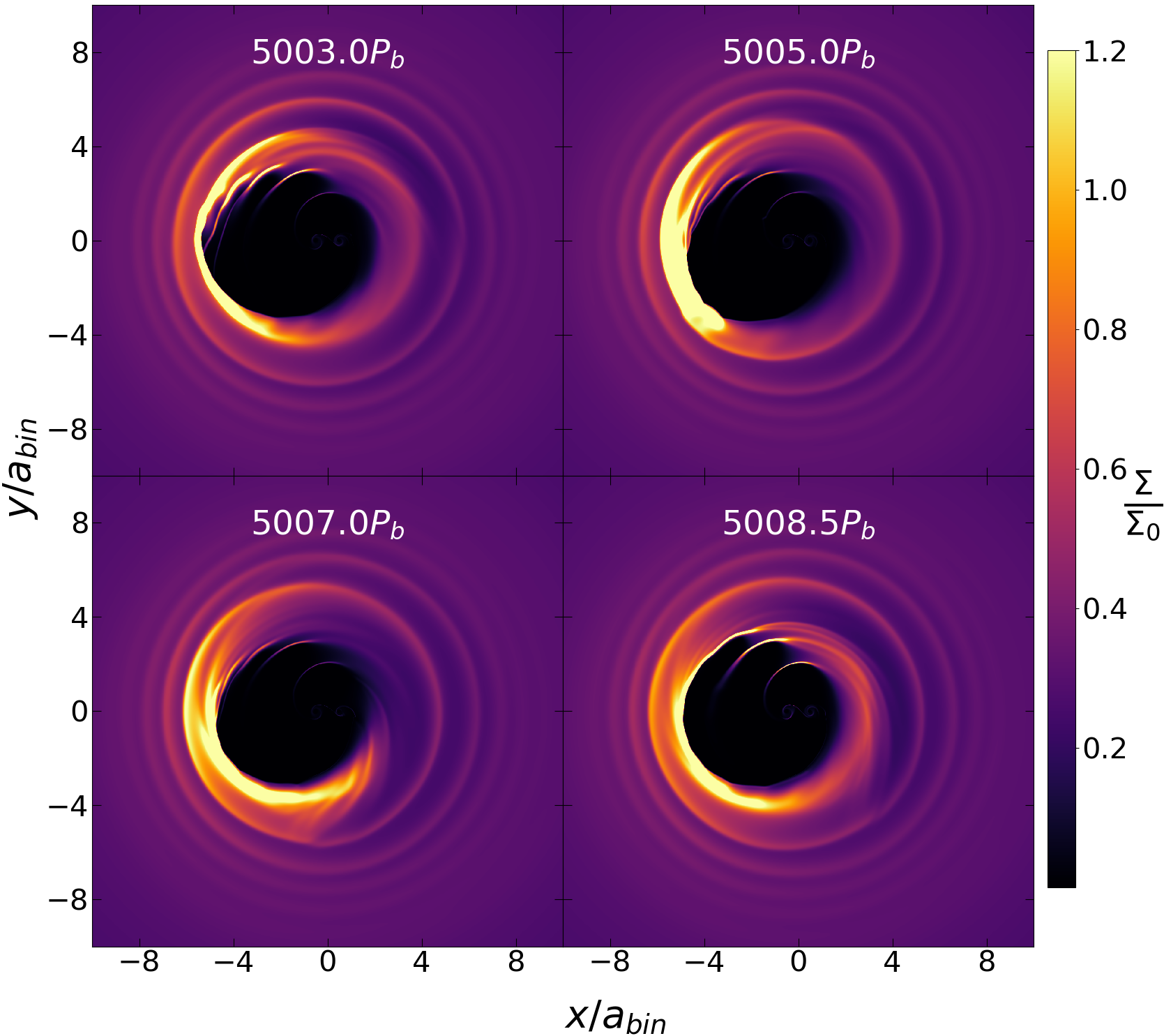}
    \caption{Surface density snapshots of the largest $\dot{m}$ viscous disc case. A density lump is created during the transition from a circular to eccentric cavity edge and orbits the cavity. It undergoes a cycle of constant creation (top left panel) and destruction (bottom left panel), with some material from the breakup being directed down the accretion streamers (bottom right panel)}
    \label{surface_den_lump}
\end{figure*}
    
    The region of the disc that produces the strongest negative torque on the binaries, and which therefore is most responsible for causing binary orbits to shrink, is known to be the cavity region \citep{Tiede2020}. Circular binaries are expected to carve out an eccentric cavity with a semi-major axis of $a \approx 2.5a_{\rm b}$ \citep{Artymowicz1994}. We therefore calculate the $\ell_0$ contribution from the gas outside the ballistic zone up to a radius of $r = 2.5a_{\rm b}$. One must be careful not to think of this region as containing the entire cavity, as the eccentricity of the disc causes the cavity wall to extend interior and exterior to a circle of radius $r = 2.5a_{\rm b}$. It is clear from Fig.~\ref{l0_all_contribution} that this part of the disc is indeed responsible for the strongest negative torque on the binaries for all simulations.

    Figure \ref{l0_all_contribution} shows the value of $\ell_0$ in the outer regions of the disc, $r>2.5a_{\rm b}$. However, as previously mentioned, the outer edge of the cavity will extend into this circular region. Gravitational torque maps of the disc show that the torque from the gas that orbits significantly outside of the  vicinity of the cavity wall (i.e. $r \gtrsim 7.0a_{\rm b}$) cancels out and so makes very little contribution to $\ell_0$. Thus, it is appropriate to combine the cavity and outer disc regions and analyse their collective effect on the orbital evolution, instead of focusing separately on what we call the cavity and outer regions. It is immediately clear that this combination contributes most heavily to changes in $\ell_0$ across the whole suite of simulations and dictates the shape of the relationship between $\ell_0$ and $\dot{m}$ in Fig.~\ref{l0_all}.

    Since the combination of the cavity and outer regions corresponds to all of the gas outside of radius $a_{\rm b}$, centred on the centre of mass, and that this combination is the most dominant factor responsible for changes to $\ell_0$, understanding the morphology of the disc is key to understanding the influence of this region, and this is considered below in Section~\ref{sect:disc_morphology}.

    To clarify which effects are responsible for causing the binary orbits to shrink, and to understand why the wind-driven discs are predicted to drive more rapid orbital shrinkage than the equivalent viscous discs, we consider eqn~\ref{eq:adot} and plot the different contributions to this expression in Fig.~\ref{fig:adot-contributions}. The top panel shows the torque contributions to $\dot{a}/a$ from each region of the disc described above for both the viscous and wind-driven runs, and here the accretion contribution includes only the accretion of momentum. The bottom panel shows the total torque contribution, the contribution from the mass accretion rate onto the binaries and the sum of these. Note the units associated with the quantities plotted in Fig.~\ref{fig:adot-contributions} are dimensionless code units, as here we are only interested in the relative contributions to $\dot a/a$.

    Looking at the top panel of Fig.~\ref{fig:adot-contributions}, we see the inner regions containing the mini-discs provide a positive torque contribution to $\dot a/a$, and that the torque in the viscous case is always significantly larger than in the case of the wind-driven discs. This could be attributed to the fact that the mini-discs in the wind-driven simulations are slightly smaller and less dense than in the viscous case, in part because these models are less effective at driving gas into the cavity region compared to their viscous counterparts (see the discussion below about the cavity structures and Fig.~\ref{1d_dens}), and in part because the wind-driven torque simply removes angular momentum from the mini-discs, and hence causes them to shrink in radius. Viscosity, however, redistributes the angular momentum and causes the mini-discs to expand to their tidal truncation radius. We see also that the accretion of momentum provides a positive torque contribution, and again this torque is larger in the viscous models because of the somewhat higher accretion rates. The outer disc and cavity regions always provide a negative torque contribution, and we see that the viscous cases again provide torques that are larger in magnitude than the wind-driven cases. 

    Turning now to the lower panel of Fig.~\ref{fig:adot-contributions}, we see that in the viscous cases the total torque contribution to $\dot a/a$ is actually positive in all runs, and it is only the contribution from the mass accretion onto the binaries that ultimately causes the orbits to shrink. In the wind-driven cases, however, both the total torque contribution and the mass accretion act to shrink the binary orbit, explaining why the simulations presented here predict that the in-spiral of binaries should occur more rapidly in wind-driven discs when compared to equivalent viscous discs.

\section{Disc morphology evolution} \label{sect:disc_morphology}

\begin{figure}
    \centering
    \includegraphics[width = 8.25cm, height = 5.5cm]{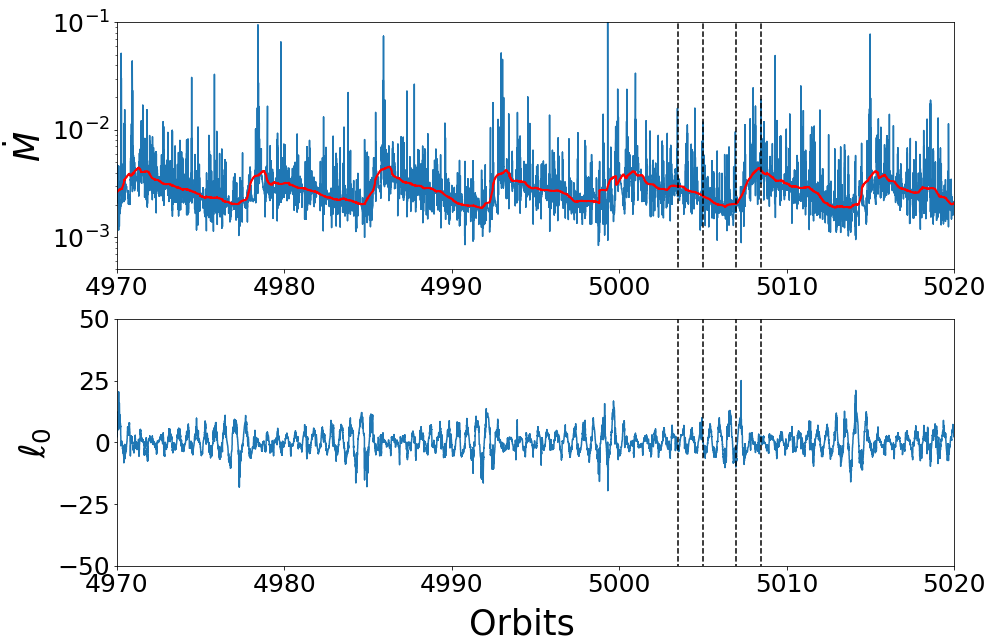}
    \caption{Mass accretion rate (top) and accretion eigenvalue (bottom) for the largest $\dot{m}$ viscous run highlighting the effect of the density lump on the measured parameters. The vertical dashed lines correspond to panels in Figure \ref{surface_den_lump}.}
    \label{mdot_l0_50_orbits}
\end{figure}

\begin{figure}
    \centering
    \includegraphics[width = 8.25cm, height = 5.5cm]{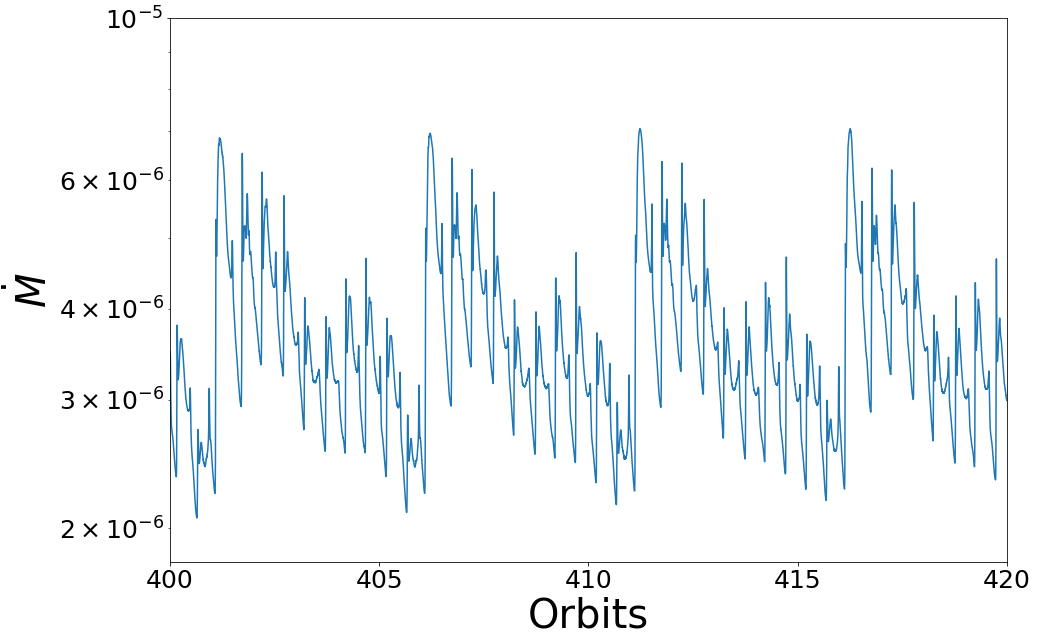}
    \caption{Mass accretion rate for the Mach 10 case ($h = 0.1, \alpha = 0.1$) in Section \ref{sect:tiede_comparison}. This highlights the lump-modulated accretion variability of period 5 orbits, equivalent to previous findings \citep[e.g.][]{Miranda2017, Dittmann2022, Wang2022}.}
    \label{fig:mdot_h}
\end{figure}

    We now examine the morphology of the disc in more detail and study how it influences the mass accretion rate, accretion eigenvalue and migration behaviour.
    
    Since the surface density is initialised with a power law profile without an already depleted cavity region, the gravitational torque due to the binary is solely responsible for clearing the cavity. This clearing occurs during the 500 binary orbit mass ramping phase, where the binary mass ratio increases from 0 to 1. When the stars have reached their final masses, in contrast to Figs.~\ref{all_alpha_den} and \ref{all_wind_den}, the cavity edge is circular and not eccentric. A transition to an eccentric cavity edge must therefore occur sometime later in the simulation. For the highest $\dot{m}$ wind-driven case, this transition is seen to occur after between 2000 and 3000 binary orbits, as shown in Fig.~\ref{surface_den_ctoe}. The transition from a circular to eccentric cavity edge is accompanied by significant changes in the mass accretion rate and accretion eigenvalue, as seen in Figs.~\ref{wind_mdot} and \ref{wind_l0}, respectively. This behaviour is observed across all simulations except for the largest $\dot{m}$ viscous run. In this model the cavity edge becomes eccentric during the initial 500 orbit mass ramping period when the binary is first introduced. The change in disc structure therefore marks the end of the initial transient phase and a quasi-steady state in the measured parameters is reached shortly after. 

    During the transition from a circular to an eccentric cavity edge, a density lump is also formed during all of the viscous and wind-driven simulations. To highlight this feature, Fig.~\ref{surface_den_lump} shows snapshots of the surface density for the largest $\dot{m}$ viscous case covering a time interval of 8 orbits. This lump is observed to orbit around the edge of the cavity with the same period as the local gas, and for our Mach 20 discs the orbital period of the lump is approximately 7 binary orbit periods. This feature has also been observed and commented upon extensively in previous works \citep[e.g.][]{Miranda2017, Dittmann2022, Wang2022}. At a certain orbital phase, the spiral density waves excited by the binary appear to congregate and impact the same region of the cavity wall. This creates an accumulation of gas at this position and leads to formation of a density lump. The lump then orbits around the cavity edge until it is broken up as it approaches the pericentre of the eccentric cavity. Some of the gas associated with the dispersing lump is directed down the accretion streamers towards the binary components, whilst what is left rejoins the over-dense region at the apocentre as the next density lump is being formed. In all runs this cycle starts as soon as the disc becomes eccentric and continues for the rest of the simulation. The lump becomes less prominent for lower mass flux rates with there also being a reduction in size and density when moving from viscous to inviscid wind-driven discs.

    The orbiting density lump has an effect on the short term variability of the mass accretion rate and accretion eigenvalue. As the lump is most prominent in the largest $\dot{m}$ viscous case, we choose this simulation for further analysis. Figure~\ref{mdot_l0_50_orbits} shows the mass accretion rate and the accretion eigenvalue over a period of 50 binary orbits. The red line in the mass accretion panel is an orbit averaged value, shown in order to reduce the noise and highlight meaningful changes in the mass accretion rate, and the times associated with the vertical dashed lines correspond to the surface density snapshots in Fig.~\ref{surface_den_lump}. As mentioned above, when the density lump is broken up some gas is directed down the accretion streamers towards the binary components. This can be seen in the mass accretion rate (especially in the averaged value indicated by the red line), where a sharp increase in the amount of mass being accreted occurs just after the density lump has directed material down the accretion streamers. The mass accretion rate then steadily decreases over a period of $\sim$7 binary orbits until the density lump directs mass down the streamers again. We note this is has been observed in many previous works \citep[e.g.][]{Miranda2017, Dittmann2022, Wang2022}, and is commonly referred to as the lump-modulated accretion variability. This modulation has typically been shown to have a period of 5 binary orbits, however these works have used larger values for both the viscosity ($\alpha = 0.1$) and disc aspect ratio ($h = 0.1$). Fig. \ref{fig:mdot_h} plots the mass accretion rate for the Mach 10 run with $\alpha = 0.1$ from Section \ref{sect:tiede_comparison}, showing a modulation period of 5 binary orbits and therefore agreeing with previous works. This is to be contrasted with the 7 binary orbit periods associated with the density lump in the Mach 20 disc models. Clearly the orbital period of the lump depends on the structure of the CBD and hence on the disc parameters.
    
    Looking to the bottom panel of Fig.~\ref{mdot_l0_50_orbits}, we notice an inverted behaviour in the accretion eigenvalue when we look at the upper envelope of the (not averaged) value of $\ell_0$.  Instead of sharply increasing and steadily declining, the envelope increases and suddenly drops (the higher frequency oscillations in this panel are due to the binary's orientation with respect to the eccentric disc, resulting in two oscillations in $\ell_0$ per orbit). As the density lump approaches the pericentre, by definition the lump is closer to the stellar components and therefore produces a larger torque on them. The lump is then broken up, dispersing its gas content over a larger area and reducing the torque on the binary. Since the accretion eigenvalue is defined as $\ell_0 = \dot{J}_{\rm b}/\dot{M}_{\rm b}$, material being directed down the accretion streams onto the binaries during the breakup results in a further decrease in the value of $\ell_0$, and explains why the magnitude of the accretion eigenvalue is the inverse of the mass accretion rate, at least when we look at the envelope associated with the time evolution of $\ell_0$.

\begin{figure}
    \centering
    \includegraphics[width = 8.25cm, height = 5.5cm]{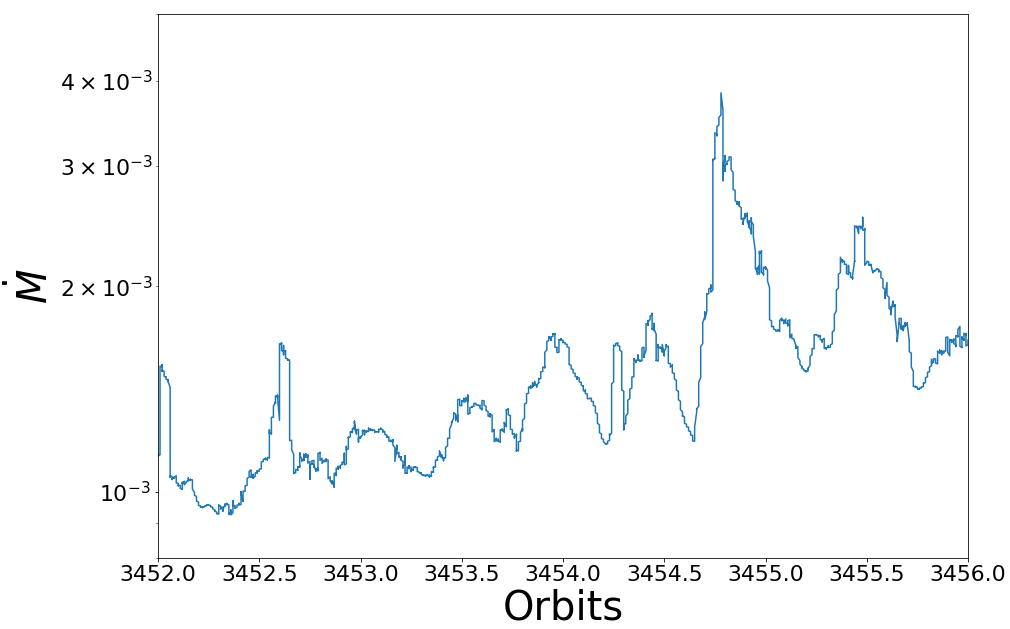}
    \caption{Binary mass accretion rate shown over a shorter time interval for the largest $\dot{m}$ wind-driven run, in order to illustrate the fact that the accretion rate varies with a period equal to half the binary orbital period.}
    \label{fig:mdot-short-timescale}
\end{figure}

\begin{figure}
    \centering
    \includegraphics[width = 8cm, height = 8cm]{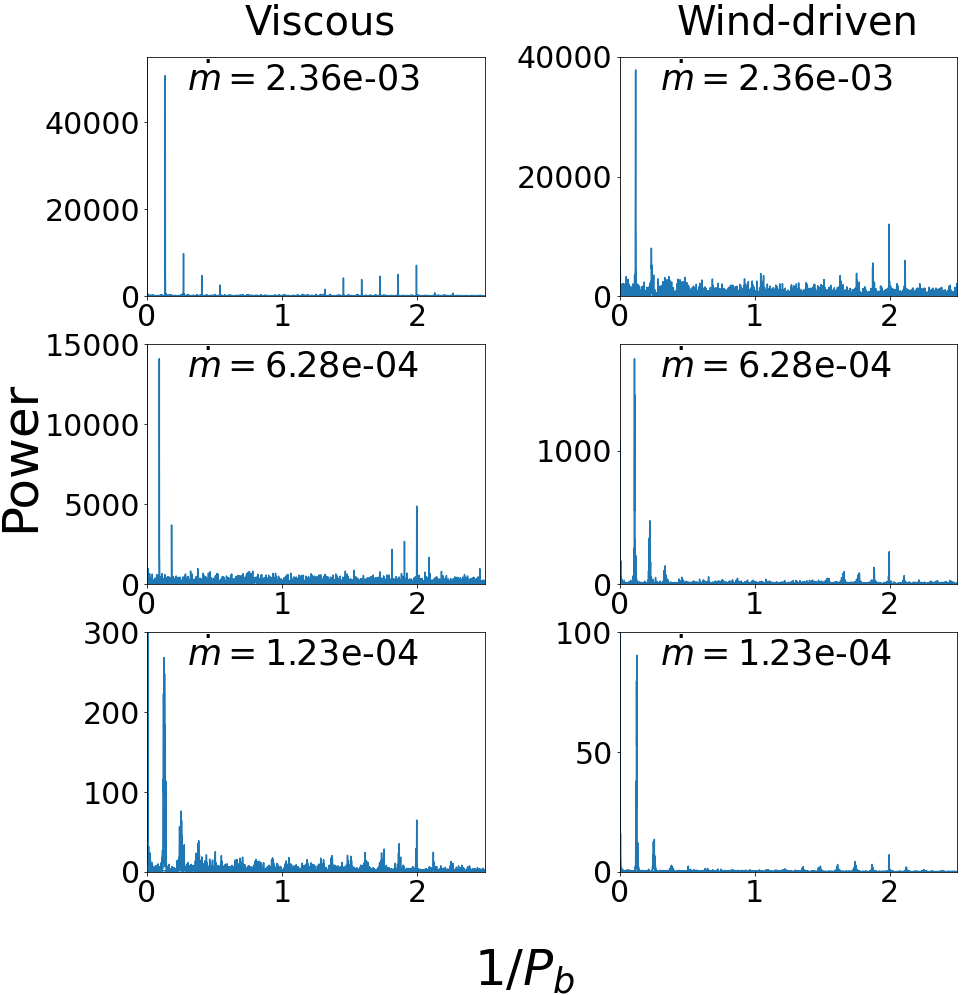}
    \caption{Power Spectrum of the mass accretion. There is a clear peak at 2 indicating accretion variation due to the motion of the binary and a large peak at 1/7 from the modulation due to the lump.}
    \label{fig:mdot_power_spectrum}
\end{figure}

    The orbit averaged mass accretion rate increases by around 30-40\% due to the break up of the orbiting density lump, however the instantaneously measured mass accretion rate can vary by as much as two orders of magnitude. This raises the question of whether this dramatic variation in accretion rate could be observed in the light curves of young binary systems that have circular orbits as considered here. Furthermore, from simple geometric considerations, we would expect an equal mass binary on a circular orbit within the eccentric cavity of a CBD to show variation in the accretion rate with a period equal to half the binary orbit, since this is the time between successive conjunctions between one of the binary components and the pericentre of the disc, which is approximately when accretion streamers are pulled from the cavity wall. In the discussion above, we noted that the time variation of $\ell_0$ showed a distinctive signal at half the binary orbit period, primarily because the torque due to the cavity-plus-outer regions of the CBD varies on this timescale. However, closer inspection of the accretion rate on short time scales also shows the expected signal at half the binary orbit period, as illustrated by Fig.~\ref{fig:mdot-short-timescale}. To investigate further, we Fourier analysed the accretion rate time series for both viscous and wind-driven disc models, and the resulting power spectra are shown in Fig.~\ref{fig:mdot_power_spectrum}. As expected, we see a strong signal at low frequency corresponding to $\sim 7$ binary orbit periods because of the effect of the density lump. In addition, however, we also see a prominent signal popping up at a higher frequency equal to 2, corresponding to half the binary orbit period. We note that \citet{Muñoz2016} report accretion rates onto the individual components from their simulation of a binary on a circular orbit (see their Fig.~3), and they note there is a half period lag between the accretion rate onto each star reaching repeated maxima, which corresponds to the half orbital period modulation of the total accretion rate we have discussed above.

\begin{figure}
    \centering
    \includegraphics[width = 8cm, height = 6cm]{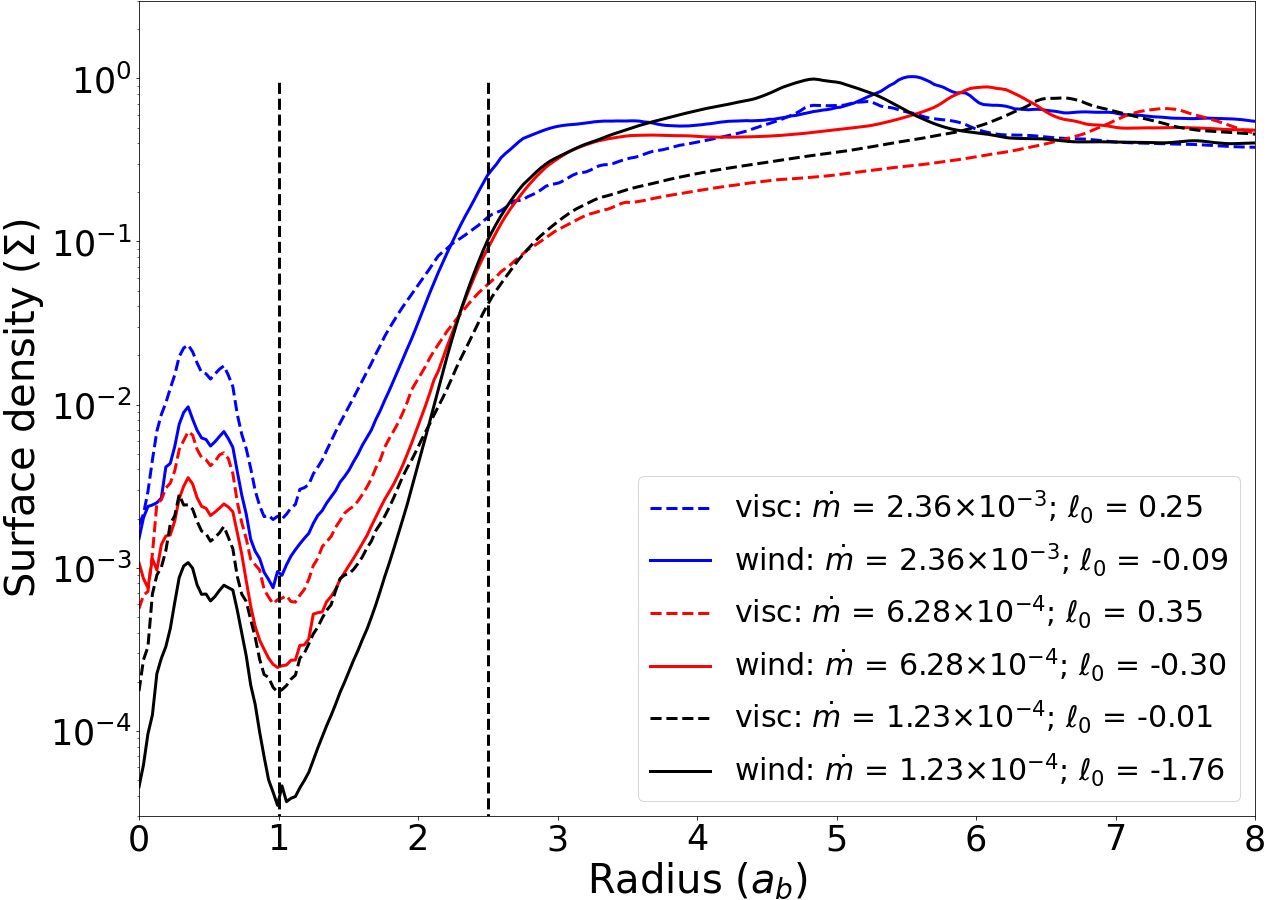}
    \caption{1D surface density profiles. The vertical dashed lines at a radius of $r = 1$ and $2.5a_{\rm{b}}$ approximately represents the outer edge of the circumstellar discs and the inner edge of the circumbinary disc respectively.}
    \label{1d_dens}
\end{figure}

    This modulation of the accretion rate is one phenomenon that could allow comparison between observations of pre-main sequence binaries on circular orbits and the simulations presented here. The well-known systems AK Sco, UZ Tau and DQ Tau are on quite eccentric orbits \citep[see table 3 in][]{Czekala2019}, and so cannot be compared with our simulations. V4046 Sgr, however, consists of nearly equal mass components on circular orbits, and evidence of variation in H$\alpha$ emission occurring with a period equal to half of the binary orbital period has been reported by \citet{Quast2000}, consistent with the time variation of the accretion rate obtained in our simulations. So far, however, no variation on the longer time scale of $\sim 7$ binary orbits associated with growth and break-up of the density lump have been reported for observations of V4046 Sgr, and identification of such a signal in observational data would provide confirmation that such a feature does indeed arise in close binary systems surrounded by CBDs. Identification of the precise period associated with this feature would also provide strong constraints on the physical properties of the disc in the cavity region, since we have shown that the orbital period of the lump depends on the disc thickness/Mach number.
    
    Finally, in Fig.~\ref{1d_dens} we show azimuthally averaged surface density profiles which are also averaged in time from surface density snapshots with intervals of 100 binary orbits over the final 1000 binary orbits of each simulation. These demonstrate that the cavity has a steeper outer edge and is deeper for the wind-driven models, as expected because the diffusive behaviour of the viscous models is absent in the wind-driven models. This figure also shows the expected behaviour as a function of the mass flux through the disc, with the lowest $\dot{m}$ runs showing the deepest cavities. The shapes of cavity outer edges and the depths of cavities near the binaries are consistent with expectations about the torque contributions to $\dot a/a$ shown in Fig.~\ref{fig:adot-contributions}, where it was noted that the torques in the viscous discs were larger in magnitude than in the corresponding wind-driven discs.

\section{Summary and conclusions} \label{sect:summary_concl}

    We have presented a suite of simulations that examine the interaction between a central binary system and a circumbinary disc, with the primary aim of determining how the interaction changes when angular momentum is transported in the disc by viscosity compared to when it is transported by an external torque arising from a magnetised wind. In this initial proof-of-concept study, we consider equal-mass binary components on fixed circular orbits, and 2D disc models defined on a Cartesian refined-mesh where the magnetised wind torque is mimicked by an external torque prescription. Our main focus is on pre-main sequence binaries, although our results may also have applicability to the evolution of binary black holes.

    We begin by calibrating the numerical viscosity, and find that at a fiducial radius of $r=1$ the numerical viscosity in our production runs corresponds to $\alpha \sim 10^{-5}$. Hence it is too small to have a significant impact on the results of our simulations.

    We next calibrate the code by running circumbinary disc simulations as a function of the Mach number (or equivalently the aspect ratio) in the disc. We obtain results that are in good agreement with previous studies \citep{Muñoz2019,Tiede2020,Penzlin2022}, with the disc developing an inner cavity that is surrounded by a precessing, eccentric circumbinary disc. Furthermore, we find that there is a transition in behaviour when increasing the Mach number above $\sim 20$, such that binary orbits are predicted to shrink rather than expand due to interaction with the surrounding gas \citep{Tiede2020,Penzlin2022}.

    The main result of this paper is that for disc models with Mach number ${\cal M}=20$ ($h=0.05$), simulations that mimic wind-driven accretion through the disc predict that binary orbits should shrink faster compared to equivalent viscous disc models (i.e. those with the same global mass flux through the disc). This behaviour occurs primarily because the positive torque experienced by the binary due to mini-discs surrounding each component is relatively large in the viscous models, slowing the in-spiral rate. Using reasonable estimates for the masses of the binary systems and CBDs that we simulate, the estimated in-spiral times for pre-main sequence binaries on circular orbits with initial separations of $\sim 1$~AU are in the range $\sim 10^6$--$10^7$~yr, indicating that orbital shrinkage through interaction with CBDs of initially well separated binary systems is a viable mechanism for forming short period binaries.

    Significant variations in the mass accretion rates and the values of $\ell_0$ are observed on short and intermediate timescales for all runs. These arise because of the orbital phasing of the binary components relative to the eccentric precessing disc, and because all runs show the development of a density lump that corotates with disc material at the edge of the cavity once the disc becomes noticeably eccentric, in agreement with previous work \citep{Miranda2017, Dittmann2022, Wang2022}. The break up of this lump, as it approaches the pericentre of the disc cavity edge, leads to additional gas being pulled into accretion streamers that feed the mini-discs around the individual stars, causing modulation of the accretion rate.  The close pre-main sequence binary system V4046 Sgr consists of nearly equal mass components on circular orbits \citep[e.g.][]{Czekala2019}, and hence is a good system to compare with our simulations. Although no variation in accretion rate has been reported that corresponds to the evolution of a density lump, there is evidence of variation in H$\alpha$ emission occurring with a period equal to half of the binary orbital period \citep{Quast2000}, consistent with the time variation of the accretion rate obtained in our simulations due to the binary orbiting within a slowly precessing eccentric cavity.

    In spite of the simplified treatment of magnetic effects in this proof-of-concept study, it nonetheless lays the groundwork for future simulations that examine the orbital evolution of pre-main sequence binaries due to interactions with their circumbinary discs. In particular, it will be important to examine the evolution of binary orbits using live binaries, since the interaction between disc and binary is bound to change as the orbit itself evolves, especially if the orbital eccentricity grows. In addition, it will also be important to move towards much more sophisticated models of wind-driven discs that actually simulate explicitly the evolution of the magnetic field and its effect on the disc. Recent work that examines the evolution of gas giant planets embedded in magnetised discs shows that the magnetic field and its torque behaves differently near the edge of the tidally-induced gap compared to torque prescriptions similar to that used in this paper \citep{Lega2022,Nelson2023,Aoyama2023,Wafflard-Fernandez2023}, and an open question is how the magnetic field and its associated torque will evolve in a circumbinary disc? These and other issues will be the subject of future work.

  \section*{Acknowledgements}
 RPN is supported by STFC grants ST/P000592/1, ST/X000931/1 and by the Leverhulme Trust through grant RPG-2018-418. GAT is supported by an STFC PhD studentship. This research utilized Queen Mary's Apocrita HPC facility, supported by QMUL Research-IT (http://doi.org/10.5281/zenodo.438045). This work was performed using the DiRAC Data Intensive service at Leicester, operated by the University of Leicester IT Services, which forms part of the STFC DiRAC HPC Facility (www.dirac.ac.uk). The equipment was funded by BEIS capital funding via STFC capital grants ST/K000373/1 and ST/R002363/1 and STFC DiRAC Operations grant ST/R001014/1. DiRAC is part of the National e-Infrastructure. All plots in this paper were made with the Python library \texttt{matplotlib} \citep{Hunter2007}.

\section*{Data Availability}

Data from our numerical models are available upon reasonable request to the corresponding author.

%
%

\bibliographystyle{mnras}
\bibliography{bibliography}

\newpage
\appendix

\section{Ballistic zone tests}
\label{Sec:Appendix2}
\begin{figure}
    \centering
    \includegraphics[scale = 0.25]{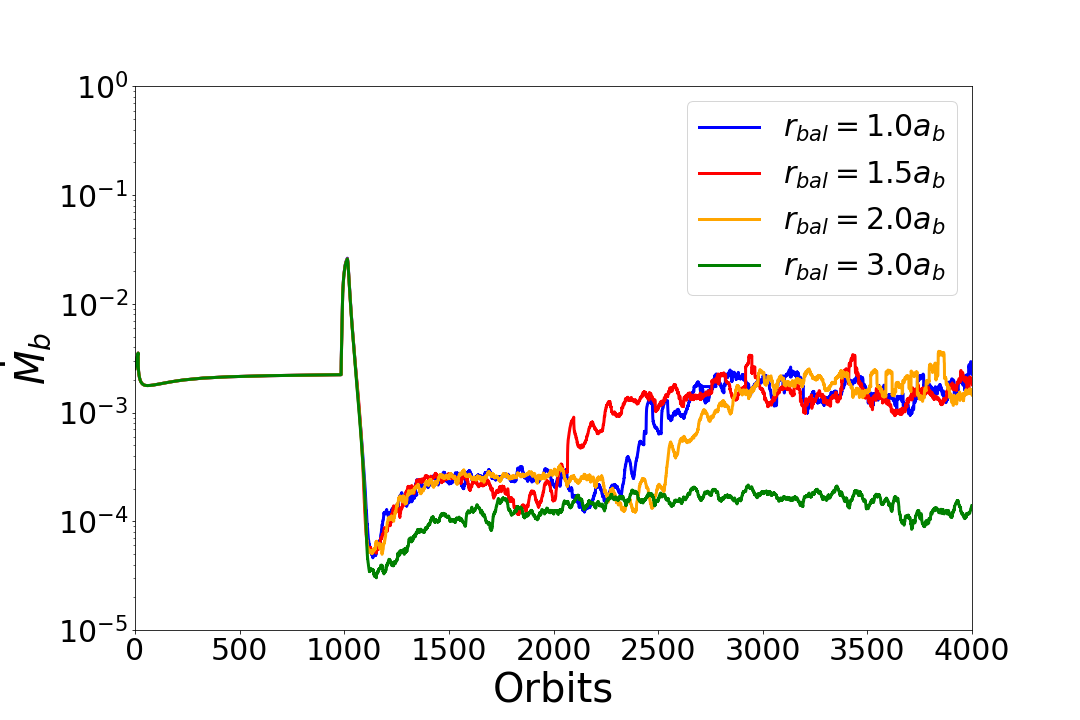}
    \caption{30-orbit averaged mass accretion rate for ballistic zone radii of $r_\textup{bal} = [1.0,1.5,2.0,3.0]a_\textup{b}$.}
    \label{fig:ballistic_mdot}
\end{figure}

\begin{figure}
    \centering
    \includegraphics[scale = 0.25]{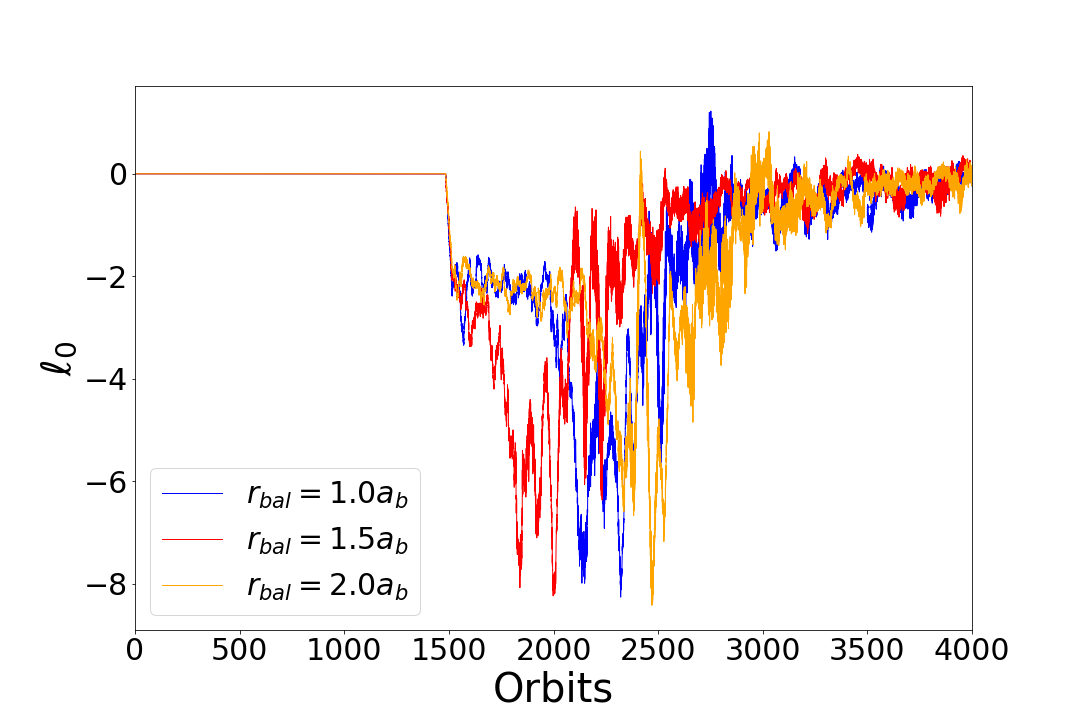}
    \caption{30-orbit accretion eigenvalue for different ballistic zone radii. The ballistic zone with radius $3a_\textup{b}$ is not plotted as the $\ell_{0}$ variability is very large (between -20 and 10) throughout the full 4000 orbits.}
    \label{fig:ballistic_l0}
\end{figure}

Since the ballistic zone we define for the wind driven disc cases is not a particularly physical representation of the torque experienced by a wind-driven circumbinary disc, we varied the radius where the wind torque is set to 0 in order to test if it has an effect on the evolution of the system. We ran this test on the largest $\dot{m} = 2.35\times 10^{-3}$ case for 4000 binary orbits and choose $r_\textup{bal} = [1.5, 2.0, 3.0] a_\textup{b}$. We plot the 30-orbit averaged mass accretion rate and accretion eigenvalue in Figs.~\ref{fig:ballistic_mdot} and \ref{fig:ballistic_l0}. These figures also include the first 4000 orbits of the result from Section~\ref{sect:wind_results}. The averaged accretion eigenvalue between orbits 3750 and 4000 have been calculated to be $\ell_0 = [-0.13, -0.11, -0.14]$ for ballistic radii of $r_\textup{bal} = [1.0, 1.5, 2.0]a_\textup{b}$ respectively. So although increasing the ballistic zone radius also increases the time to reach a quasi-steady state, both mass accretion and $\ell_0$ show little change in the converged values between $r_\textup{bal}$ = 1.0, 1.5 and 2.0 $a_\textup{b}$. When the ballistic zone has a radius of $r_\textup{bal} = 3.0$, an eccentric cavity does not form within the simulation time, and the mass accretion rate and $\ell_0$ do not reach the same final values. We do not plot $\ell_0$ for this case as the fluctuations are so great (between -20 and 10) that it obscures the rest of the plot. From this test we conclude that as long as the size of the ballistic zone is not too large, i.e.. $>3.0a_\textup{b}$, the results of the simulations are reasonably well converged.

\section{Angular momentum conservation for viscous ring}
\label{Sec:Appendix1}
\begin{figure}
    \centering
    \includegraphics[scale = 0.25]{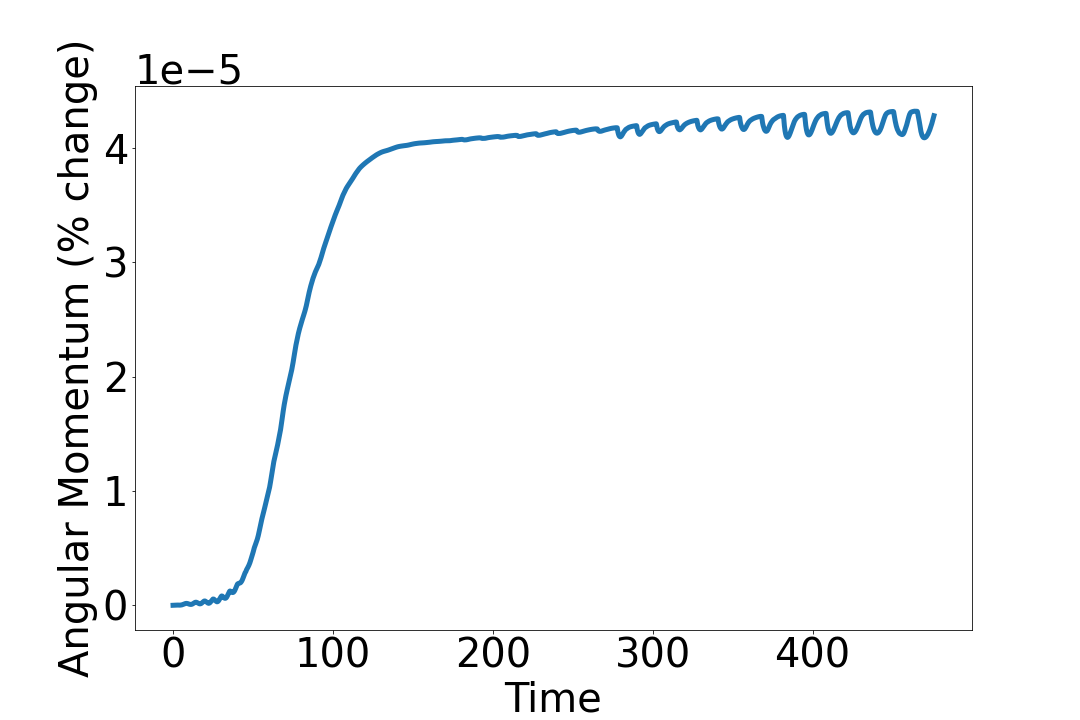}
    \caption{Percentage change in the total angular momentum measured between radii $r=0.2$-$1.8$ during a simulation of the viscous ring problem. Time is shown in units of $\Omega^{-1}$ at $r=1$.}
    \label{fig:visc_ring_am}
\end{figure}

The use of a cartesian grid in the simulations means that the angular momentum of an orbiting disc or ring is not expected to be conserved to machine precision. Here we examine the global angular momentum conservation during a simulation of the viscous ring problem described in Section~\ref{sect:numerical_visc}. We use a resolution of $2400 \times 2400$ cells. The angular momentum, $\mathbf{J}= \mathbf{r} \times \mathbf{p}$, where $\mathbf{p}$ is the linear momentum, is calculated for each cell between the radii 0.2 and 1.8 and is then summed over all cells. Figure~\ref{fig:visc_ring_am} plots the percentage change in the total angular momentum relative to the initial conditions, and we note that a very small change of $4 \times 10^{-5}\%$ occurs over the duration simulation, which ran for $\sim 450$~$\Omega^{-1}$, where $\Omega$ is the angular velocity measured at a radius of unity. The observed increase in angular momentum up until $t\approx 100$ is probably due to residual pressure effects causing the ring to expand slightly, since the time over which it occurs is approximately equivalent to the ring width divided by the sound speed. However, once this phase is over the angular momentum is very well conserved.

\bsp	
\label{lastpage}
\end{document}